\documentclass[12pt]{article}
\usepackage{amsmath, amsthm, amssymb, amsfonts,float}
\usepackage[font=small, labelfont=bf]{caption}
\usepackage[caption=false]{subfig}
\usepackage{jheppub}
\usepackage{amsmath, amsthm, amssymb, amsfonts}
\usepackage{comment}
\usepackage{graphicx}
\usepackage{float}
\usepackage[numbers,sort&compress]{natbib}
\usepackage{framed}
\usepackage{physics}
\usepackage{multirow,array}

\usepackage[capitalise]{cleveref}

\usepackage{booktabs}
\usepackage{verbatim}

\newcommand{\rp}{r_p}
\newcommand{\ystar}{\hat{y}_\star}

\newcommand{\roughly}[1]{\mathrel{\raise.3ex\hbox{$#1$\kern-0.85em\lower1ex\hbox{$\sim$}}}}
\newcommand{\lsim}{\roughly<}
\newcommand{\gsim}{\roughly>}

\def\cA{{\cal A}}

\def\cE{{\cal E}}

\def\cK{{\cal K}}
\def\cL{{\cal L}}
\def\cM{{\cal M}}

\def\cO{{\cal O}}
\def\cP{{\cal P}}

\newbox\charbox
\newbox\slabox
\def\slsh#1{{      
        \setbox\charbox=\hbox{$#1$}
        \setbox\slabox=\hbox{$/$}
        \dimen\charbox=\ht\slabox
        \advance\dimen\charbox by -\dp\slabox
        \advance\dimen\charbox by -\ht\charbox
        \advance\dimen\charbox by \dp\charbox
        \divide\dimen\charbox by 2
        \raise-\dimen\charbox\hbox to \wd\charbox{\hss/\hss}
        \llap{$#1$}
}}

\def\exd{{\hbox{d}}}

\def\bfx{{\bf x}}

\def\bfa{{\bf a}}

\def\cA{{\cal A}}

\def\cE{{\cal E}}
\def\cI{{\cal I}}

\def\cL{{\cal L}}
\def\cM{{\cal M}}
\def\cO{{\cal O}}

\def\cP{{\cal P}}

\def\Rc#1{R_{{#1}}}

\def\bea{\begin{eqnarray}}
\def\eea{\end{eqnarray}}
\def\be{\begin{equation}}
\def\ee{\end{equation}}

\def\ssE{{\scriptscriptstyle E}}

\def\ssI{{\scriptscriptstyle I}}

\def\ssR{{\scriptscriptstyle R}}

\def\nn{\nonumber}

\def\({\left(}
\def\){\right)}

\def\pref#1{(\ref{#1})}
\def\eff{{\rm eff}}

\def\rp{r_p}
\def\Rw{R_w}
\def\Rc{R_c}
\def\e{\textrm{e}}
\def\iu{\textrm{i}}
\def\epscut{\varepsilon}
\def\modE{\mathfrak{E}}
\def\argE{\mathfrak{e}}
\def\pol{\kappa }
\def\modA{\mathfrak{A}}
\def\argA{\mathfrak{a}}
\def\sgn{\textrm{sgn}}
\def\permfs{\epsilon_0}
\def\OmegaD{\Omega_{d-1}}

\def\qDens{\varrho}
\newcommand{\hlambda}{\hat{\lambda}}
\newcommand{\hLambda}{\hat{\Lambda}}
\newcommand{\hX}{\hlambda_\ssR}
\newcommand{\hY}{\hlambda_\ssI}

\newcommand{\citer}[1]{Ref. \cite{#1}}
\makeatletter
\newcommand\footnoteref[1]{\protected@xdef\@thefnmark{\ref{#1}}\@footnotemark}
\makeatother

\graphicspath{{Figures/}}

\title{Fall to the Centre in Atom Traps\\ and Point-Particle EFT for Absorptive Systems}

\author[a,b]{R. Plestid,}
\author[a,b]{C.P. Burgess}
\author[a]{and D.H.J. O'Dell}
\affiliation[a]{Department of Physics \& Astronomy, McMaster University\\ $\phantom{wd}$ 1280 Main Street West, Hamilton Ontario, Canada L8S 4M1}
\vspace{2mm}
\affiliation[b]{Perimeter Institute for Theoretical Physics\\  $\phantom{wd}$ 31 Caroline Street North, Waterloo Ontario, Canada  N2L 2Y5 }
\vspace{2mm}

\emailAdd{cburgess@perimeterinstitute.ca, plestird@mcmaster.ca, dodell@mcmaster.ca}

\date{\today}

\abstract {Polarizable atoms interacting with a charged wire do so through an inverse-square potential, $V = - g/r^2$. This system is known to realize scale invariance in a nontrivial way and to be subject to ambiguities associated with the choice of boundary condition at the origin, often termed the problem of `fall to the center'. Point-particle effective field theory (PPEFT) provides a systematic framework for determining the boundary condition in terms of the properties of the source residing at the origin. We apply this formalism to the charged-wire/polarizable-atom problem, finding a result that is not a self-adjoint extension because of absorption of atoms by the wire. We explore the RG flow of the complex coupling constant for the dominant low-energy effective interactions, finding flows whose character is qualitatively different when $g$ is above or below a critical value, $g_c$. Unlike the self-adjoint case, (complex) fixed points exist when $g> g_c$, which we show correspond to perfect absorber (or perfect emitter) boundary conditions. We describe experimental consequences for wire-atom interactions and the possibility of observing the anomalous breaking of scale invariance. 
}

\begin{document}
\maketitle
\section{Introduction}

The quantum mechanics of the attractive inverse-square potential poses conceptual challenges that are not encountered for quantum motion in potentials that are less singular at the origin  \cite{Landau1982}. The main characteristic feature is the need to specify a boundary condition at the origin that cannot simply be the default boundary condition of boundedness at $r = 0$ used in less singular situations (such as the Coulomb potential).   

The default boundary condition cannot be used because of competition between the inverse-square potential and the centrifugal barrier that means that for some values of system parameters {\em both} linearly independent solutions to the radial Schr\"odinger equation are singular at $r = 0$. Because both solutions are singular the condition of boundedness at the origin does not provide a useful criterion for distinguishing amongst them, and so some other choice is needed.


The heart of the problem is that, by itself, the inverse-square potential does not lead to a completely specified physical problem and extra ingredients must be added in order to fully determine it. The literature is full of proposals for boundary conditions that might be appropriate \cite{Kase1950, Perelomov1970, Alliluev1972, Jackiw1991,Gupta:1993id, Beane2000, Coon2002, Bawin2003, Mueller2004, Braaten2004, Werner2008, Bouaziz2014}. These are not unique and many are guided by the requirement that the boundary condition preserve the self-adjointness of the system's hamiltonian (the so-called self-adjoint extensions). When this is so, the new boundary condition can be regarded as a choice for the scattering phase associated with whatever physics happens to be sitting at the origin. 

On the one hand, it is clear that the inverse square potential only describes the long-range (low-energy) part of the problem and the `right' boundary condition in a particular physical situation should depend on the properties of whatever the underlying object sitting at the origin turns out to be. On the other hand, it is also clear that complete knowledge of {\em all} of the microscopic details of this object should not be necessary if the object is sufficiently small. What is missing is some sort of systematic algorithm that quantifies precisely which source properties govern the boundary condition of fields near the origin; a kind of `generalized multipole expansion', but as applied to more general fields --- such as the Schr\"odinger field, for example --- than just electrostatics.

\citer{Burgess2016} proposes an algorithm for identifying what these generalized multipoles might be for the Schr\"odinger field, and how to infer from them the relevant boundary condition at the origin. (Refs.~\cite{Burgess2016a} and \cite{Burgess:2017ekj} do the same for the Klein-Gordon and Dirac fields, respectively.) The proposal is to determine the boundary condition directly from an action, $S_p$, that describes how the first-quantized source at the origin couples to the fields of interest. This action then describes all of the local interactions with the source, which can then be converted into a boundary condition on these fields at the origin. For instance, for a Schr\"odinger field coupled to a static compact object situated at $\bfx = 0$ in flat spacetime we have 
\be \label{Spintro}
   S_p = \int \exd^4x \; \delta^d(\bfx) \, \cL_p(\Psi, \Psi^*) \,,
\ee
where $d$ is the co-dimension of the source and $\cL_p = -h \, \Psi^* \Psi + \cdots$, where $h$ is a coupling constant that parameterizes the source-field interaction and the ellipses represent terms with more powers of $\Psi$, $\Psi^*$ and their derivatives. 

It is the effective couplings like $h$ in this lagrangian that play the role of generalized multipole moments\footnote{Indeed for electromagnetic fields, $A_\mu$, with $\cL_p$ restricted to terms linear in $A_\mu$, these couplings literally are the usual multipole moments.} of the source, in the sense that only the lowest-dimension interactions are relevant in the limit where the source's size, $\rp$, is much smaller than the other scales of interest. That is, the order of magnitude of the coefficient of any particular operator in $S_p$ is generically set by the dimensionally appropriate power of the source's small size, $\rp$, making terms with more powers of $\Psi$ and/or derivatives parametrically sub-dominant as $\rp \to 0$. Provided one is only interested in physics extending over a scale $a$ much larger than $\rp$, one can expand observables in powers of $\rp/a$, and only the lowest-dimension interactions in $S_p$ can contribute to any fixed and low order in this expansion. 

The connection to boundary conditions comes because once $S_p$ is known the boundary condition for the bulk field is found by a kind of generalized Gauss' law wherein the near-source limit of $\OmegaD r^{d-2} \partial_r \Psi$  is equated with $\delta S_p/\delta \Psi^*$, where $\OmegaD$ is the area of the unit $(d-1)$-sphere. (The boundary condition is slightly different for fermions \cite{Burgess:2017ekj}). The resulting construction has a definite effective-Lagrangian flavour \cite{Weinberg:1978kz} (for a review, see for instance \cite{Burgess:2007pt}), with the only difference being that the relevant Lagrangian is first-quantized from the point of view of the source, which could be a single domain wall, string-like defect or point particle. For this reason this boundary condition proposal is known as `point-particle effective field theory' (PPEFT).

In this paper we extend the discussion of \cite{Burgess2016} to objects that are either sources or sinks of probability for the degrees of freedom of interest. Such sources are of interest in describing many systems, ranging from wires or other localized objects interacting with non-relativistic atoms in a trap through to quantum fields interacting with black holes \cite{Goldberger2005}. We find, as in the case of probability-conserving sources, that the effective couplings of the action must be renormalized because fields like $\Psi$ typically diverge at the position of a source. As a consequence they evolve under an RG-flow, whose form we characterize in the complex-coupling plane. Physical quantities must be invariant under this flow and so can only be functions of RG-invariants, whose form we identify and classify. We explicitly relate these invariants to observables such as the (complex-valued) scattering phase shift or bound-state energies and, in the presence of absorption, decay lifetimes. We recover as a special case the conventional absorptive cross-section\footnote{Defined by the class of trajectories that fall to the center in finite time. These have an impact parameter satisfying $b<b_c$, and the cross section is defined as the volume of a $d-1$ dimensional ball of radius $b_c$ (i.e.\ $\sigma_\text{abs} = \pi b_c^2$ for $d=3$) \cite{Landau1982}.} of a perfect absorber in the classical limit, in agreement with earlier uses of non-self adjoint boundary conditions for singular potentials \cite{Vogt1954,Perelomov1970,Alliluev1972}.

The remainder of the paper is organized as follows. \S\ref{sec:PPEFT} starts by reviewing how PPEFT works for the inverse-square potential, but including the possibility of probability non-conservation at the source.  \S\ref{sec:charged-wire} then introduces the concrete example of a gas of polarizable atoms interacting with a charged wire and provides explicit formulae for parameters in our PPEFT in terms of physical quantities such as applied voltages, and atomic polarizabilities. In \S\ref{sec:renormalization} we discuss the classical renormalization required when evaluating the field $\Psi$ at the source, where it typically diverges. We track the two qualitatively different kinds of renormalization-group (RG) flow, that differ according to whether or not the coupling $g$ of the attractive inverse-square potential, $V = - g/r^2$, is stronger than a critical value, $g_c$. Finally, a convenient RG-invariant parameterization of the flow is presented.  \S\ref{sec:RG-obs} then computes how observables depend on the properties of the source and in particular how they depend on the RG invariants identified in the previous section. The observables studied include both the elastic and absorptive scattering cross sections, as well as the energy eigenvalue for bound states, and their decay lifetimes in the presence of point-like absorptive physics. Particular attention is paid to providing an explicit example in which RG invariants can be extracted from experimental data in the context of the charged wire setup introduced in \S\ref{sec:charged-wire}. Next in \S\ref{sec:expt-prot} we discuss which regions of parameter space can be realistically probed in the lab, and briefly speculate on the possibility of future experiments. Finally \S\ref{sec:summary} briefly summarizes our results.

\section{PPEFT of the inverse-square potential}
\label{sec:PPEFT}

This section describes the EFT appropriate to non-relativistic atoms interacting with a source through an inverse-square potential in $d$ dimensions. We write results explicitly for the two cases of most practical interest: $d=2$ (for the charged wire) and $d = 3$ (for point-particle sources). 
\subsection{The `bulk'}
The first choice to be made is identifying the degrees of freedom whose coupling to the source is of interest. With a view to applications to atomic systems we take these to consist of non-relativistic and polarizable atoms represented by the 2nd-quantized Schr\"odinger field $\Psi$, describing (say) trapped atoms and their long-range inverse-square interaction with the source. 
\subsubsection{Action and field equations}
These atomic degrees of freedom--- hereafter called the `bulk' --- are then described by the action\footnote{In principle the terms explicitly written are also accompanied by a succession of other effective interactions, such as $g_4 |\Psi|^4$ and so on, whose couplings have dimension involving higher powers of time or length than those written (and so are relatively unimportant for low-energy, long-distance applications).}  
\begin{equation}
  \label{bulkaction}
  S_B=\int \dd t \,\dd^d x \,
   \left[ \frac{\iu}{2} \qty(\Psi^*\partial_t\Psi-\Psi\partial_t\Psi^* )
  - \frac{1}{2m}\abs{\nabla\Psi}^2+\frac{g}{{\abs{\vb{x}}^2}} |\Psi|^2 \right] \,,
\end{equation}
where $m$ is the atomic mass and $g$ is a positive real bulk-coupling parameter with dimension (in units with $\hbar = 1$) of (energy) $\times$ (length)${}^2$ that for $d=2$ is of order $\pol \qDens^2$, where $\qDens = Q/L$ is the wire's charge per unit length and $\pol$ the atom's polarizability. Physically, the inverse square potential emerges since the atoms' induced dipole moment is proportional to $\kappa \vb{E}$, and so their energy scales as $\kappa \vb{E}^2\propto 1/\vb{r}^2$. Eq.~\pref{bulkaction} has the Schr\"odinger equation as its field equation,
\be \label{SchrEq}
  \iu \, \partial_t \Psi = - \frac{1}{2m} \nabla^2 \Psi + V(\bfx) \Psi 
\ee
with $V(\bfx) = -g/|\bfx|^2$.

We note that \pref{bulkaction} and \pref{SchrEq} display an anomalous symmetry in the form of a continuous scale invariance, which is a well known feature of the inverse square potential \cite{Jackiw1991,Coon2002,Kaplan2009,Moroz2009}. This can be seen by noting that under $\bfx\rightarrow  s\bfx$ the entire right-hand side of \pref{SchrEq} scales as $1/s^2$, which can be reabsorbed via an appropriate redefinition of time. As a result, if we find a solution at one scale $\bfx$, then we naively expect there to be a continuous family of solutions with the same shape at \emph{all} scales. It turns out that this expectation is only true for specific choices of boundary conditions, with a generic choice leading to an anomalous breaking of scale invariance, the consequences of which we return to in \S\ref{sec:RG-obs}.

For the applications of interest the field $\Psi$ is expanded in a basis of mode functions, $\psi_n(x) \propto \langle 0 | \Psi (x)| n \rangle$, for a complete set of single-atom states, $|n\rangle$, so
\begin{equation}
  \Psi(x) =\sum_{n}  \bfa_{n} \, \psi_{n}(x) \,,
\end{equation}
for annihilation operators, $\bfa_n$. Since our later focus is on spherically symmetric sources these modes can be further decomposed into partial waves by writing $x = \{r, \Sigma \}$ (with $\Sigma$ the solid-angle on the $d-1$ sphere), and $n = \{s,\ell,\mu\}$ with
\begin{equation}
  \psi_n(r, \Sigma)  = \psi_{s\ell}(r) \, Y_{\ell\mu}(\Sigma) \,
\end{equation}
where $Y_{\ell {\mu}}$ are the appropriate hyper-spherical harmonics, satisfying $\nabla^2 Y_{\ell{\mu}}= - \varpi_d(\ell)Y_{\ell{\mu}}$, defined on the unit hyper-sphere labelled by $\Sigma$. 

For general $d$ the projection quantum number, $\mu$, collectively denotes a set of labels, but is simpler for the cases of practical interest. Specifically, for $d = 3$ we have $\ell = 0,1,2,\cdots$ and $\mu$ is the magnetic quantum number $\mu = -\ell, -\ell+1,\cdots,\ell -1, \ell$ while $\varpi_3 = \ell(\ell+1)$. For $d = 2$, on the other hand, there is no label $\mu$ and $\varpi_2 = \ell^2$ for $\ell = 0,\pm 1, \pm 2, \cdots$.  With these choices the equation satisfied by the radial mode functions is
\begin{equation}
  r^2\dv[2]{\psi_{s \ell}}{r}+r(d-1)\dv{\psi_{s\ell}}{r}
  - \Bigl[ \varpi_d(\ell)-2mg-2mEr^2 \Bigr] \psi_{s\ell}=0 \,,
  \label{eq:radial-equation}
\end{equation}
where $E$ is the state's energy. 

\subsubsection{Boundary condition ambiguity}

The interesting (and well-known) observation about such inverse-square potentials --- and indeed any attractive potential which satisfies $\lim_{r\rightarrow 0} \abs{V(r)}\geq \order{r^{-2}}$ --- is that they are singular enough at the origin to compete with the centrifugal barrier. This allows sufficient probability to accumulate near $r = 0$ to necessitate more careful specification of the physics of the source. More explicitly, the need to do so arises because the radial solutions in the small-$r$ limit asymptote to $\psi_\ell(r) \propto r^s$ where $s(s+d-2) = \varpi_d(\ell) - 2mg$, and so 
\be \label{spmdef}
 s = s_\pm := \frac12\left( 2-d \pm \zeta \right) \,,
\ee
with
\begin{equation} \label{zetadef}
  \zeta:=\sqrt{(d-2)^2+4[\varpi_d(\ell)-2mg]} \,.
\end{equation}
In what follows we choose $\zeta$ to be the root for which either $\zeta \ge 0$ (if $\zeta$ is real) or $\xi := -\iu \zeta \ge 0$ (if $\zeta$ is imaginary). For reasons that will be made clear in \S\ref{sec:fixed-points}, we refer to the case of $\zeta$ real as sub-critical, and to the case of $\zeta=\iu \xi$ (with $\xi$ real) as super-critical.

As mentioned earlier, this asymptotic form reveals a need for a physical criterion for choosing the boundary condition to be imposed as $r \to 0$. Although the boundary condition is often assumed to be the requirement that $\psi$ not diverge in this limit, the inverse-square potential exposes this as inadequate because (for example) whenever $2mg > \varpi_d(\ell)$ the real parts of $s_+$ and $s_-$ have the same sign. When this is true both solutions are singular (or both are not) at the origin, and so boundedness cannot be the criterion with which to ensure the eigenvalue problem is well-posed. This is particularly striking when $\zeta$ is imaginary --- as happens when $g > g_c$ with $8mg_c = (d-2)^2 + 4\varpi_d(\ell)$ --- since in this case the real parts of $s_\pm$ are identical. 

The need to choose a boundary condition near the origin is widely discussed in the literature, but the `right' choice is largely left a matter of guesswork. The choice of boundary condition often seems essentially arbitrary and many choices for dealing with it --- self-adjoint extensions, regulator potentials, and so on \cite{Kase1950, Perelomov1970, Alliluev1972, Beane2000, Coon2002, Bawin2003, Braaten2004, Werner2008, Bouaziz2014}--- are given in the literature. 

\subsection{PPEFT for the source}

As argued in \cite{Burgess2016}, and as summarized in the introduction, a very efficient way to specify the missing boundary condition uses the first-quantized effective action, $S_p$, for the source, written as a function of the `bulk' fields of interest (in our case $\Psi$). This section describes the source action relevant for a Schr\"odinger field interacting through an inverse-square potential with a localized source. With the example of a charged wire in mind we allow probability not to be conserved at the source, so the boundary condition found need not be self-adjoint.\footnote{Non-self-adjoint boundary conditions are also considered in \cite{Perelomov1970}.}   

At low energies the dominant couplings involve the lowest-dimension interactions in $S_p$ that depend on $\Psi$.  Like the bulk action, the point-particle action must respect the underlying symmetries of the problem, which in the present instance include invariance under rephasings $\Psi \to \e^{\iu \theta} \Psi$ (to ensure conservation of the number of atoms), as well as rotational symmetry. 

The lowest--dimension source interaction is then given by
\begin{equation}
  S_p=-\int \dd t \,\dd^dx \;  \delta^{(d)}(x) \Bigl[ h \abs{\Psi}^2+ \cdots \Bigr] \,,
  \label{eq:point-particle-action}
\end{equation}
where $h$ is an effective coupling with dimension (energy) $\times$ (length)${}^{d}$ --- and so is (energy) $\times$ (length)${}^3$ for a point charge in $d=3$ dimensions (such as in the applications discussed in \cite{Burgess2016a}), or is (energy) $\times$ (length)${}^2$ for the charged wire (for which $d=2$). Allowing non--unitary evolution (only at the source) means allowing the coupling $h$ in $S_p$ to be complex-valued.

The system's total action then is $S = S_B + S_p$, and so an interaction like \pref{eq:point-particle-action} contributes to the field equation \pref{SchrEq} by modifying the potential to acquire a delta-function term,
\be
 V(\bfx) =  - \frac{g}{|\bfx|^2} + h \,\delta^d(\bfx) \,.
\ee

\subsection{Boundary conditions }

As argued in detail in \cite{Burgess2016} --- and as is normally assumed for delta-function potentials in any case --- the new contribution acts to modify the boundary condition for the field $\Psi$ near the origin, with
\be \label{eq:Spbc}
   \left( \OmegaD r^{d-1} \frac{\partial \Psi}{\partial r} \right)_{r = \epscut} = 2mh \, \Psi(\epscut) \,,
\ee
where $\OmegaD$ is the area of the unit $(d-1)$-sphere. As mentioned in the Introduction, this is the analog of Gauss' law for the Schr\"odinger field.  The boundary condition is evaluated at a small but nonzero radius $r = \epscut$ because at sufficiently small $r$ a description strictly in terms of the effective action of \pref{eq:point-particle-action} breaks down, and a more complete description of the source is required (more about which below). Eq.~\pref{eq:Spbc} fixes the parameter $h$, and so solves, for this simple example, the problem of connecting the source action, $S_p$, to the small-$r$ boundary condition of the bulk field $\Psi$. 

When the physics near $r = 0$ does not conserve probability (such as for atoms interacting with the charged wire) we need not demand the coupling $h$ be real. In this case the leading low-energy implications of probability loss at the source are captured by Im$\,h$, in the same way that the leading unitary contributions of source physics at low energies are captured by Re$\,h$. 

To quantify the relation between Im$\,h$ and probability loss at the source we use the boundary condition \pref{eq:Spbc} to compute the radial probability flux operator at $r = \epscut$:
\be
 J_r(\epscut) = \frac{\iu}{2m} ( \Psi \partial_r \Psi^* - \Psi^* \partial_r \Psi)_{r=\epscut} = \iu \left( h^* - h \right)  \frac{| \Psi(\epscut)|^2 }{\OmegaD \epscut^{d-1}}  \,,
 \label{eq:prob-flux}
\ee
where $\OmegaD$ is the surface area of the unit $d-1$ sphere (i.e.\ $\OmegaD=4\pi$ for $d=3$). This shows that the operator that controls the net rate of probability flow out of a sphere of radius $r = \epscut$ is 
\be \label{netprobloss}
 {\cP} := \oint_{r=\epscut} J_r \; \exd \Sigma  = 2  \; |\Psi(\epscut)|^2 \, \hbox{Im} \, h \,.
\ee
Clearly, positive (negative) Im $h$ corresponds to the compact object at $\bfx = 0$ being a net probability source (sink).

\section{Charged-wire example \label{sec:charged-wire}}

A charged thin wire interacting with a gas of neutral atoms is a concrete physical example of a system exhibiting both fall to the centre \cite{Perelomov1970, Alliluev1972, Landau1977,Vogt1954} and non-unitary boundary conditions \cite{Perelomov1970, Alliluev1972,Goldberger:2005cd}. Furthermore, these non-unitary boundary conditions arise due to the presence of a small compact object which interacts with (and ultimately absorbs) atoms. Besides lending itself to a conceptual discussion of PPEFT, this system has the added benefit of having an explicit experimental realization \cite{Denschlag1998}. Motivated by this work, we consider a gas of neutral atoms in a grounded cylinder of radius $\Rc$ at whose center lies a thin wire of radius $\Rw \ll \Rc$. As discussed in the previous section, the influence of this wire on the atoms is decomposed into two parts: the bulk action includes a long-range background inverse-square potential, while the point-particle action parameterizes the local atom-wire coupling responsible for the absorption of atoms.

\subsection{Parameter matching}
Holding the central wire at a fixed voltage $V_w$ induces on it a nonzero linear charge density $\qDens$, given in terms of the voltage (in SI units) by 
\begin{equation}
     \qDens= \frac{2\pi \permfs \,V_w}{\ln\qty({\Rc}/{\Rw})}
      \,.
\end{equation}
The resulting electric field is radially directed with magnitude 
\begin{equation}
     \abs{\vb{E}}= \frac{\qDens}{2\pi\permfs \,r} =  \left[ \frac{V_w}{\ln\qty({\Rc}/{\Rw})} \right] \frac{1}{r}
\end{equation}
at a radial distance $r$ from the wire.

Although the atoms in the trap are neutral, their polarizability allows their internal charge distributions to adjust to the presence of this electric field thereby inducing in them a dipole moment, and leading to an effective interaction potential between the atoms and the wire of size \cite{Pethick2002}
\begin{equation}
    U(r) = -\frac{1}{2}(4\pi\permfs \,\pol)\abs{\vb{E}}^2 = - \left[
    \frac{2\pi \permfs \, \pol \, V_w^2}{\ln^2\qty(\Rc/\Rw)}\right]  \frac{1}{r^2} \,,
\end{equation}
where $\pol$ is the atomic polarizability. Comparing to the form $U(r) = - g /r^2$ defines the `bulk-coupling'
\begin{equation}
   g = \frac{2\pi \permfs \, \pol \,  V_w^2}{\ln^2(\Rc/\Rw)}  \,.
\end{equation}

Using this in the definition \pref{zetadef} for $\zeta$, specializing to $d=2$ and, restoring $\hbar$ then gives
\begin{equation}
   \zeta(V_w,\ell)=\frac{2}{\hbar}\sqrt{L^2-2 m g(V_w)}
               = 2\sqrt{\ell^2-  \qty(\frac{V_w}{V_w^{(1)}})^2 }
\end{equation} 
where $L=\ell\hbar$ with $\ell\in \mathbb{Z}$. This expression defines a useful characteristic voltage: the value $V_w = V_w^{(1)}$ for which $\zeta(V_w^{(1)},\ell=1)$ vanishes, given explicitly by
\begin{equation}
  V_w^{(1)}=  \Bigl( 0.64  \; \hbox{Volts} \Bigr) \ln\qty(\frac{\Rc}{\Rw}) 
               \sqrt{\qty(\frac{a_0^3}{\pol})\qty(\frac{1~\text{amu}}{m})} \,
               \label{eq:vc_one}
\end{equation} 
where $a_0$ is the Bohr radius. Evidently even ordinary voltages can allow complex $\zeta$ for a broad range of $\ell$.

Adjusting the voltage of the central wire allows different values of $\zeta$ to be explored for each quantum number $\ell$. Notice in particular that, since we are in two-dimensions, $\zeta$ is always imaginary for $\ell = 0$ and is always real for sufficiently large $\ell$. Denoting by $\ell_c$ the angular momentum that satisfies $\zeta(V_w, \ell_c) = 0$ we see that $\zeta$ is imaginary for all $0 \le \ell < \ell_c$. Furthermore, $\ell_c = 0$ when $V_w = 0$ and $\zeta(V_w) \leq \zeta(V_w=0)$ is real for all fixed $\ell > \ell_c$.

The above discussion also demonstrates that for $\ell\neq 0$ there exists a finite range of voltages $0<V<V_w^{(\ell)}$ such that the effective radial potential ($U_\text{eff}(r)=(\ell^2-2mg)/r^2$  for $d=2$) is repulsive. As we will discuss later [beneath eq.\ \pref{eq:epsstar-eps-rel}], a spatially extended (i.e.\ $\epscut_\star\gg \rp$), and hence experimentally observable, exotic bound state is expected in the limit that $\zeta\rightarrow 0^+$. This limit is easily achieved in the case of the p-wave state with relatively small applied voltages. 
 
The experiment discussed in \cite{Denschlag1998}, only probes the classical physics of the inverse square potential corresponding to large angular momenta. One reason for this is that the experiment used large voltages on the order of $100$ Volts. In light of the above discussion, however, it seems reasonable that the quantum physics of this system could be probed by slowly tuning the voltage on the wire, and allowing each partial wave to slowly transition from real $\zeta(\ell)$ to imaginary $\zeta(\ell)=\iu\xi(\ell)$. In this case, only a few partial waves would dominate the absorptive cross section allowing each to be addressed individually. For $\xi\sim \order{1}$ quantum effects become important and the RG flow discussed in \S\ref{sec:renormalization} results in clear phenomenological signatures. As we discuss in \S\ref{sec:infer-inlab} these signatures can be used to extract RG invariants that characterize the parameters of the PPEFT. In \S\ref{sec:expt-prot} we discuss strategies to be used in new experiments to create tunable short-range interactions between the atoms and the source (in addition to the long-range inverse square potential) such that the RG invariants mentioned above can be tuned in the lab.

\section{Renormalization and RG flows \label{sec:renormalization}}

The boundary condition \pref{eq:Spbc} is evaluated at a nonzero radius because the effective description of \pref{eq:point-particle-action} breaks down at distances smaller than the actual size of the source, $r = \epscut \lsim \rp$.

The breakdown of EFT methods at $r = 0$ is not just a conceptual point: it can also cause extrapolations to zero size purely within the effective theory to diverge or otherwise be ill-defined --- such as the evaluation of solutions to the field equations $\Psi(r=0)$ and $\partial_r \Psi(r=0)$. 

From this point of view evaluating at nonzero $r = \epscut$ regulates the divergences at $r = 0$ arising from such an extrapolation. One might worry that the necessity to do so makes predictions within the EFT $\epscut$-dependent, and so ambiguous. This does not happen, however, because all $\epscut$-dependence naively appearing in observables is cancelled by an implicit $\epscut$-dependence hidden within the renormalization\footnote{Physically, $h$ is chosen to reproduce at low-energies the full physics of the source, and the value that is required to do so depends on the value of $\epscut$ where the boundary condition is applied.}  of couplings like $h$ --- for more details see \cite{Burgess2016}. Consequently $\epscut$ can be taken to be arbitrary, provided only that it satisfies $\rp \ll \epscut \ll a$: it must be large enough to allow the full description to be well-approximated by a generalized `multipole' expansion, yet small enough so as to be much shorter than the length scale of physical interest, $a$.

\subsection{Mode expansion}

To make the small-$r$ implications of the boundary condition \pref{eq:Spbc} explicit, we use this equation to determine the single-particle mode functions, $\psi_n(x)$, within a partial-wave expansion. Separating variables and suppressing all labels except $\ell$, \pref{eq:Spbc} implies the radial mode function satisfies
\be \label{eq:Spbcmode}
   \left( \OmegaD r^{d-1} \frac{\exd \psi_\ell}{\exd r} \right)_{r = \epscut} = 2mh \, \psi_\ell(\epscut) \,.
\ee

Writing the two linearly independent solutions of \pref{eq:radial-equation} as $\psi_{\ell\pm}(r)$ --- with the two solutions differing in their asymptotic small-$r$ power-law forms $\psi_{\ell\pm} (r)\propto r^{s_\pm}$ with $s_\pm$ given by \pref{spmdef} --- the general solution before imposing boundary conditions is  
\be \label{Psipm}
 \psi_\ell = C_{\ell+} \psi_{\ell+} + C_{\ell-} \psi_{\ell-} \,.
\ee
with (for the Schr\"odinger action) the radial mode functions given explicitly by
\begin{align}\label{psielldef}
  \psi_{\ell\pm}(kr)&=(2\iu kr)^{\frac{1}{2}(2-d\pm\zeta)}\e^{-\iu kr}
  \cM\qty[\frac{1}{2}(1\pm \zeta),1\pm \zeta; 2\iu kr]\nn\\
  &=(2\iu kr)^{\tfrac{1}{2}(d-2)}
  2^{\pm \zeta}\Gamma\qty(1\pm \tfrac{\zeta}{2})I_{\pm\zeta}(\iu kr)
\end{align}
where $\cM[a,b;z]=1+(a/b)z + ...$ is the confluent hypergeometric function, and $I_\nu(z)$ is the modified Bessel function of the first kind. 

The boundary condition \pref{eq:Spbc} fixes the ratio $C_{\ell-}/C_{\ell+}$ through the relation 
\be \label{psiellbc}
 \frac{2mh}{\OmegaD \epscut^{d-1}} = \Bigl( \partial_r \ln \psi_{\ell} \Bigr)_{r=\epscut} = \left(  \frac{C_{\ell+} \psi'_{\ell+} + C_{\ell-} \psi'_{\ell-}}{C_{\ell+} \psi_{\ell+} + C_{\ell-} \psi_{\ell-}} \right)_{r=\epscut} \,,
\ee
where $\psi_{\ell\pm}=\psi_{\ell\pm}(kr)$ with $k^2=2mE$. It is important to emphasize that this choice is physical, because the ratio $C_{\ell-}/C_{\ell+}$ is in one-to-one correspondence with physical observables. This can be seen by studying \pref{psielldef} in the large-$r$ limit, where $C_{\ell-}/C_{\ell+}$ dictates scattering phase shifts, and the energy at which a bound state is normalizable at infinity.

For instance, in the special case where $d=3$ and $g = 0$ we have $s_+= \ell > s_- = -(\ell +1)$ and so $\psi_{\ell-} \propto r^{s_-} = r^{-\ell-1} \to \infty$ as $r \to 0$ for all $\ell$, while $\psi_{\ell+} \propto r^{s_+} = r^\ell$ is bounded there.  In this case the traditional boundary condition, which assumes $\psi_\ell$ must be bounded at $r=0$, corresponds to $C_{\ell-} = 0$. This is only consistent with \pref{psiellbc} if $h \to 0$ as $\epscut \to 0$, i.e.\ in the absence of a direct coupling between $\Psi$ and the source. 

Eq.~\pref{psiellbc} is used extensively in what follows, and can be read in either of two complementary ways. First, as derived above, it shows how the ratio $C_{\ell+}/C_{\ell-}$ is determined in terms of the microscopic coupling $h$ by the boundary condition at $r = \epscut$. Alternatively, because physical observables cannot depend on the arbitrary scale $\epscut$ neither can the ratio $C_{\ell+}/C_{\ell-}$ and \pref{psiellbc} shows what the $\epscut$-dependence of $h(\epscut)$ must be in order for this to be true. In general, the presence of the source makes the external mode functions diverge as $\epscut \to 0$, and this divergence is cancelled in observables by the $\epscut$-dependent running of effective couplings like $h$. 

Note that the requirement that $h(\epscut)$ runs also means that in principle if the source couples in an $\ell$-dependent manner, then separate couplings might also be required for each $\ell$. It can be tempting to assume that only $\ell = 0$ modes couple to a localized source like \pref{eq:point-particle-action}, because `normally' $\psi_\ell(r) \propto r^\ell$ as $r \to 0$ and so $\psi(0)$ vanishes unless $\ell = 0$. However, we have seen that this argument is circular: the assumed $r$-dependence is wrong precisely when $h \ne 0$. Instead, for a generic source, nothing prevents $\psi_\ell$ from diverging at $r = 0$ for any $\ell$. Consequently it is occasionally useful to make this dependence on $\ell$ more explicit by writing \pref{eq:point-particle-action} as
\be  
  S_p=- \sum_\ell \int \dd t  \;  \Bigl[ h_\ell(\epscut) \abs{\Psi_\ell(\epscut)}^2+ \cdots \Bigr] \,,
  \label{eq:ppa-ell}
\ee
which underlines that there can be an independent coupling\footnote{This is best interpreted as a boundary action defined on the surface at $r = \epscut$, as in \cite{Burgess2016}. Notice that \pref{eq:ppa-ell} in general need not be a local expression once written in coordinate space on this surface.} $h_\ell$ for each $\ell$. 

\subsection{RG evolution}
We next make the $\epscut$-dependence of $h$ more explicit, extending the discussion of \cite{Burgess2016} to the case of complex $h$. To this end we use the approximate small-$r$ limit $\psi_{\ell\pm} (r) \simeq (2\iu kr)^{s_\pm}$ in \pref{psiellbc} to obtain the more explicit relation
\bea \label{psiellbc2}
 \frac{2mh}{\OmegaD \epscut^{d-1}}  \simeq  \frac{1}{\epscut} \left[\frac{ s_+(2\iu k \epscut)^{s_+-s_-} + (C_{\ell-}/C_{\ell+}) s_-}{(2\iu k\epscut)^{s_+-s_-} + (C_{\ell-}/C_{\ell+}) }\right] \,,
\eea
where $k^2=2mE$ and $s_\pm = \frac12(2-d\pm \zeta)$. The main assumption required to use \pref{psiellbc2} instead of \pref{psiellbc} is that subdominant terms in the small-$r$ asymptotic form for each of the $\psi_{\ell\pm} (r)$ can be dropped. It is consistent to do this while also keeping the subdominant function $\psi_{\ell-}\sim r^{s_-}$ separately from $\psi_{\ell+}\sim r^{s_+}$ provided that Re $(s_+-s_-) = \frac12 \,\textrm{Re}~\zeta<1$\footnote{We note however, that eq.\ \pref{psiellbc2} applies, at least approximately, over a surprisingly large range of $\epscut$ as discussed in \cref{app:surprise}.}. 

To read off the evolution of $h$ from this it is useful to write \pref{psiellbc2} as
\begin{equation}
  \label{prelambdadef}
  \frac{2mh}{\OmegaD\epscut^{d-1}} \simeq \frac{1}{2\epscut} \Bigl( 2-d + \hlambda \Bigr)
\end{equation}
which defines the convenient dimensionless variable $\hlambda$ by\footnote{In \cite{Burgess2016} the right-most expression of eq.\ \ref{eq:RGE2} is taken as the definition for $\hlambda$. By defining $\hlambda$ as in \cref{lambdadef}, we may study its RG exactly at the cost of a more complicated relationship between $\hlambda$ and $h$. See \cref{app:RG} for a more detailed discussion.}
\begin{equation} \label{lambdadef}
  \hlambda:= \left( \frac{1-R}{1+R} \right) \zeta 
  \,,
\end{equation}
with
\be\label{Rdef}
R(\epscut):=\frac{C_{\ell-}}{C_{\ell+}}(2\iu k\epscut)^{-\zeta} \,.
\ee

As ever, it is useful to write the evolution \pref{psiellbc2} in differential form, and this can be done by directly differentiating \pref{psiellbc2} with the ratio $C_{\ell-}/C_{\ell+}$ held fixed (such that physical observables are $\epscut$ independent), and re-expressing the result in terms of $h$ (or $\hlambda$) again using \pref{psiellbc2}, leading to \cite{Burgess2016}
\begin{equation}
  \epscut\, \frac{\exd \hlambda}{\exd \epscut}  \simeq  \frac{1}{2} \Bigl( \zeta^2-\hlambda^2 \Bigr) \,.
  \label{eq:RGE-Re}
\end{equation}
Because $\zeta$ is real or pure imaginary (for real $g$) it follows that \pref{eq:RGE-Re} preserves the reality of $\hlambda$ for all $\epscut$ if it is real at any specific value $\epscut_0$. This shows that it is RG-invariant to impose a Hermitian boundary condition.

\begin{figure}
  \centering
  \begin{subfloat}[][$\zeta$ real]%
    {\includegraphics[width=0.45\linewidth]{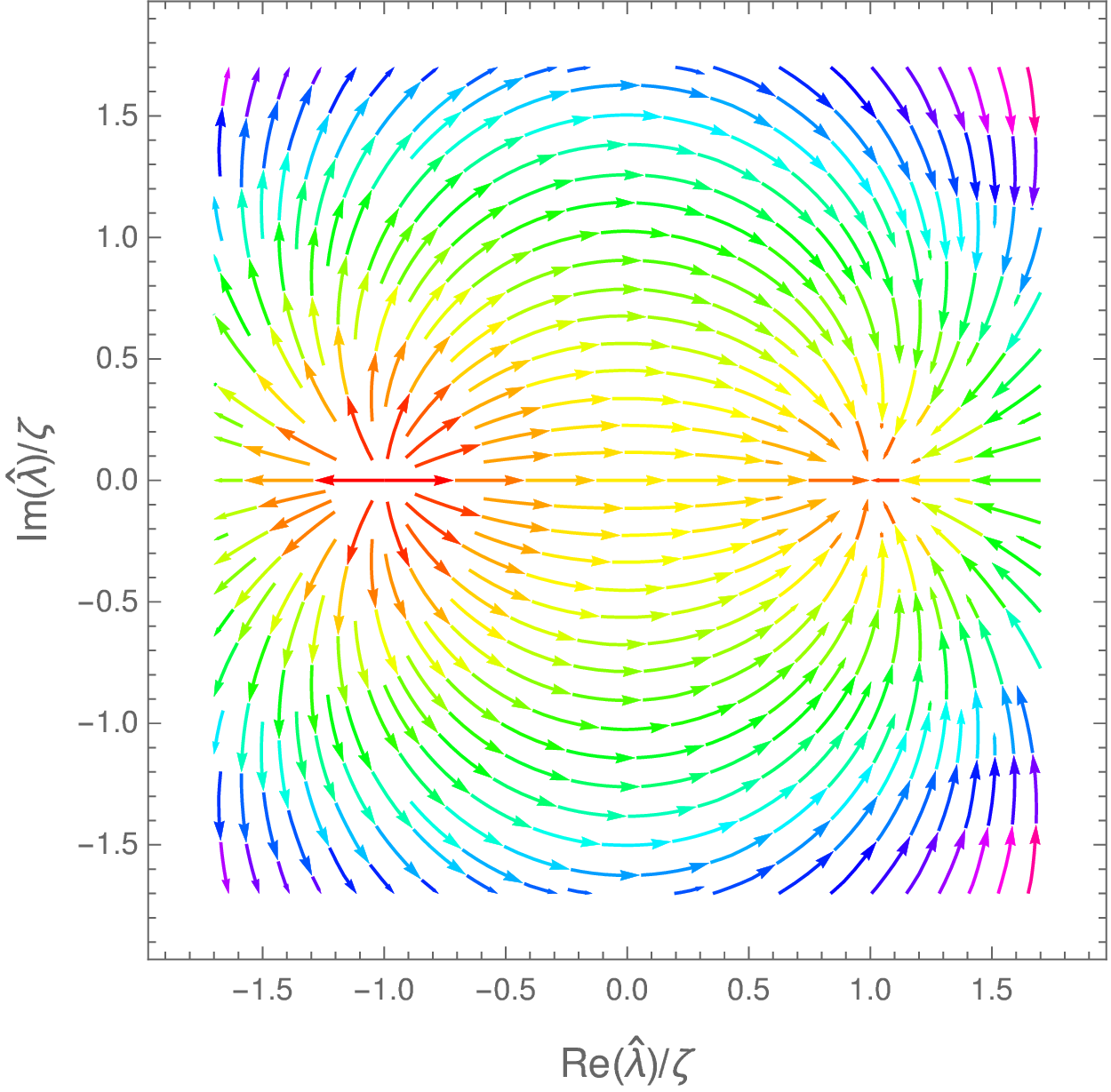}~~
      \label{fig:phase-port-a}} %
  \end{subfloat}%
  \begin{subfloat}[][$\zeta=\iu \xi$ imaginary]%
   {    \includegraphics[width=0.45\linewidth]{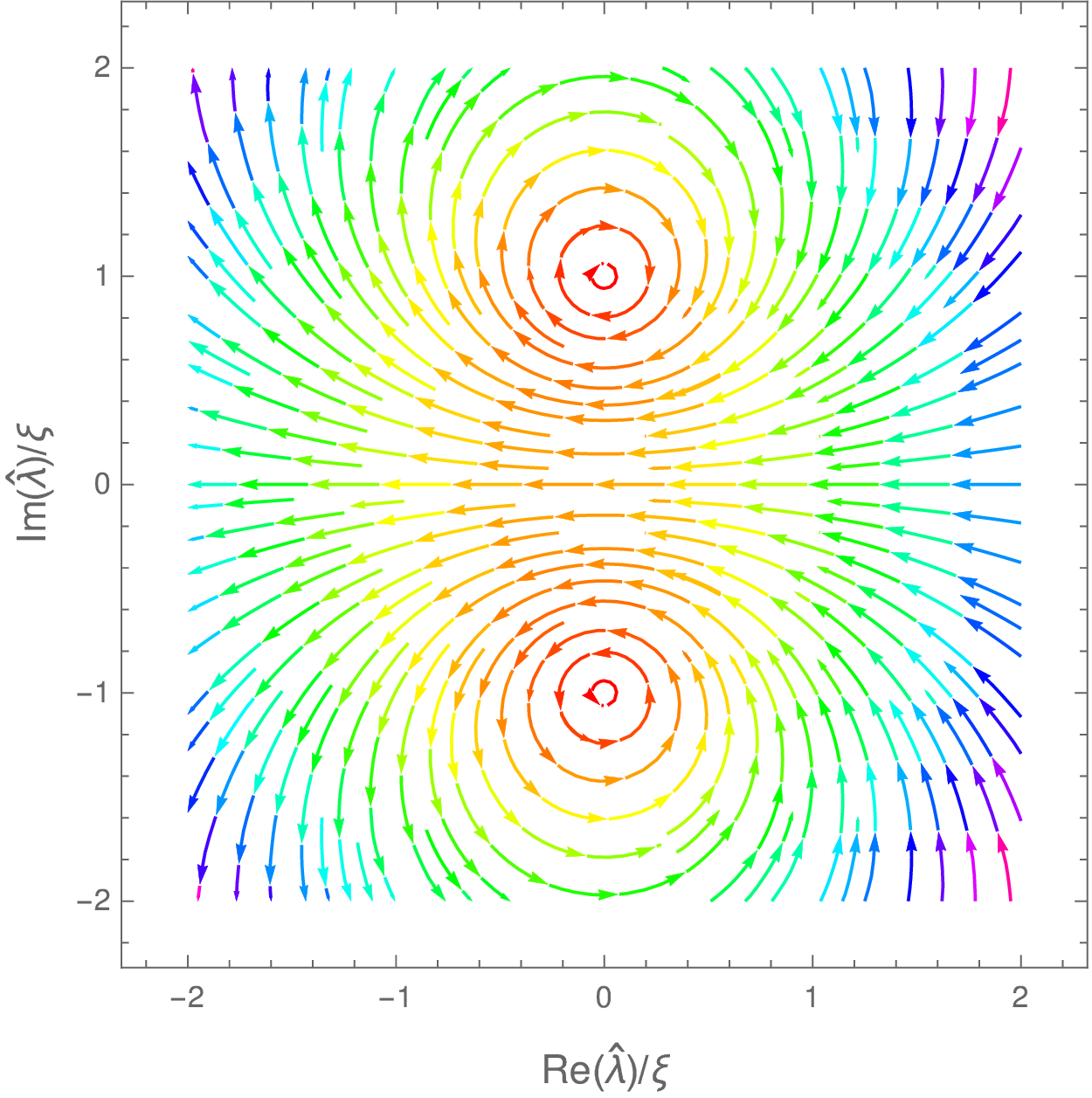}
      \label{fig:phase-port-b}} %
 \end{subfloat} 
   \caption{Phase portrait of the differential equation \pref{eq:RGE-Re} governing the RG flow of $\hlambda$ (related to $h$ the coupling constant in $S_p$ \pref{Spintro}) for $\zeta$ real \protect\subref{fig:phase-port-a} and $\zeta=\iu \xi$ imaginary \protect\subref{fig:phase-port-b}. Arrows indicate the direction of flow as $\epscut$ increases and colours indicate the speed of the flow (decreasing from violet to red). For both panels the real axis is a separatrix, so that the property of being absorptive or emissive is RG invariant dictated by the sign of $\hY$.  \label{fig:phase-portraits}}
\end{figure}

\begin{figure}
  \begin{subfloat}[][$\zeta$ real]%
    {  \includegraphics[width=0.45\linewidth]{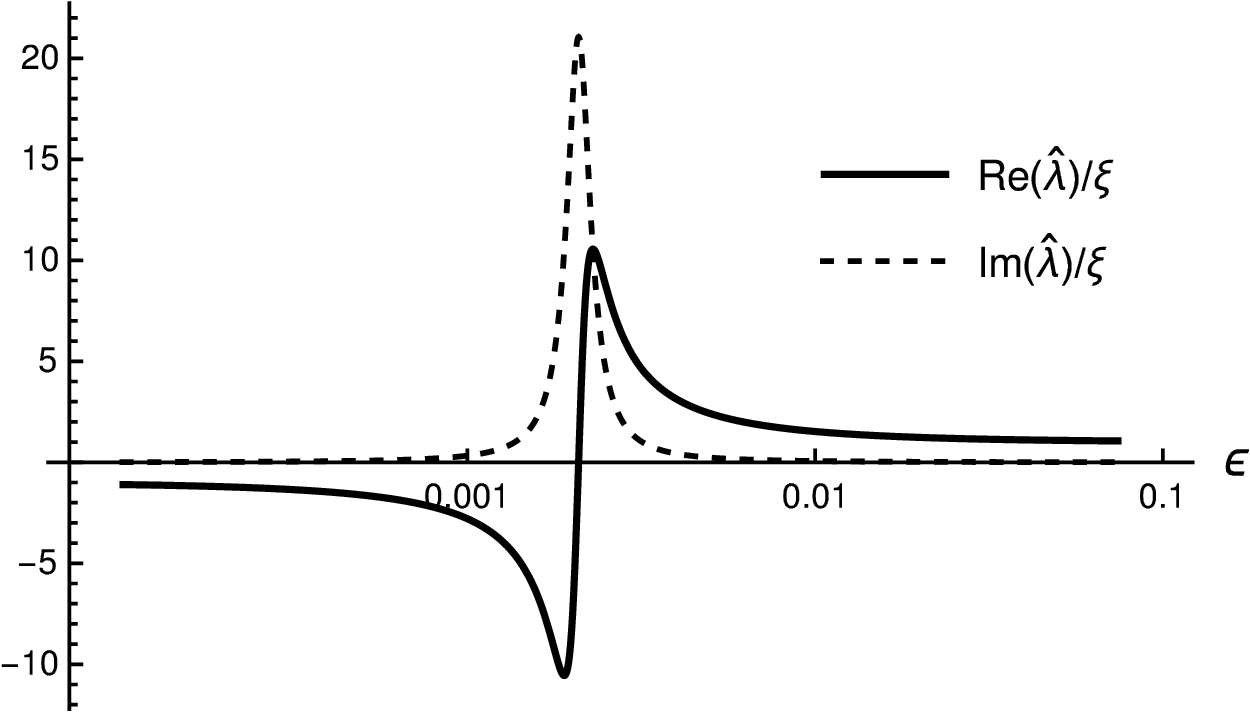}~~
      \label{fig:rep-flow-a}} %
  \end{subfloat}%
  \begin{subfloat}[][$\zeta=\iu \xi$ imaginary]%
   {    \includegraphics[width=0.45\linewidth]{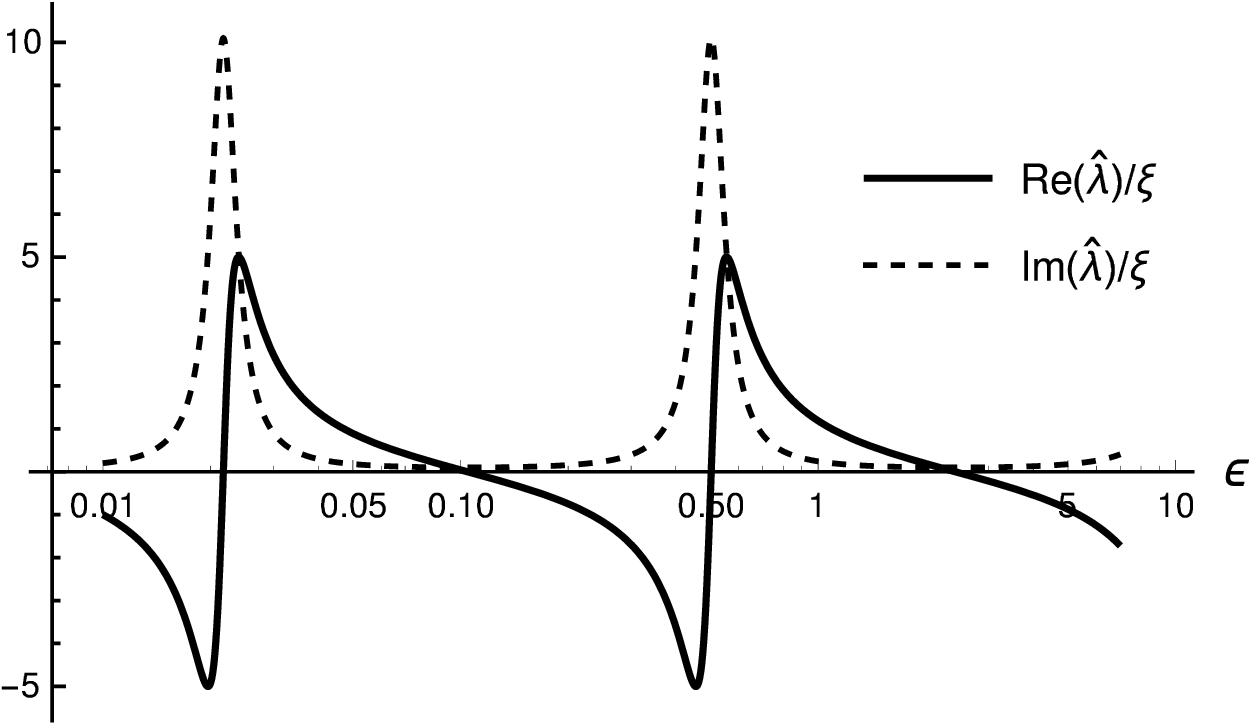}
      \label{fig:rep-flow-b}} %
 \end{subfloat} 
  \caption{Typical RG flows \pref{eq:RGE2} for $\hX$ and $\hY$ for $\zeta$ real \protect\subref{fig:rep-flow-a} and $\zeta=\iu \xi$ imaginary \protect\subref{fig:rep-flow-b}. Each RG flow picks out a special length scale $\epscut_\star$ where $\text{Re}\hlambda =0$ and so breaks the continuous scale invariance.  \label{fig:rep-flows}}
\end{figure}

Integrating \pref{eq:RGE-Re} with the initial condition $\hlambda(\epscut_0) = \hlambda_0$ gives the re-parameterization of \pref{psiellbc2} given in \cite{Burgess2016},
\begin{equation}
  \frac{\hlambda(\epscut)}{\zeta} = \frac{\hlambda_0 + \zeta \, \tanh\qty[\tfrac{\zeta}{2}   \ln(\epscut/\epscut_0)]}
         {\zeta +\hlambda_0 \tanh\qty[\tfrac{\zeta}{2}\ln(\epscut/\epscut_0)]} = \frac{(\hlambda_0 + \zeta)(\epscut/\epscut_0)^\zeta + (\hlambda_0 - \zeta)}{(\hlambda_0 + \zeta)(\epscut/\epscut_0)^\zeta - (\hlambda_0 - \zeta)} \,, 
         \label{eq:RGE2}
\end{equation}
which applies equally well for real or complex $\hlambda$. Figure \ref{fig:phase-portraits} plots this solution in the complex $\hlambda$-plane --- see also \citer{Moroz2009} --- where the left panel corresponds to $\zeta$ real and the right panel to $\zeta$ imaginary. The dependence of the real and imaginary parts, $\hX(\epscut)$ and $\hY(\epscut)$, on $\epscut$ is similarly given in Figure \ref{fig:rep-flows}. 

 

Once a numerical value for $\hlambda$ is specified at a specific $\epscut = \epscut_0$ the integration constants associated with \pref{eq:RGE-Re} are obtained by solving \pref{psiellbc}, leading to 
\begin{equation}
  \frac{C_-}{C_+}= 
  \frac{\zeta-\hlambda(\epscut)}{\zeta+\hlambda(\epscut)} \;
   \qty(2k\epscut)^\zeta \e^{\iu \pi\zeta/2},
  \label{eq:boundary-ratio}
\end{equation}
where to avoid clutter we henceforth suppress the index $\ell$. The point of \pref{psiellbc2} -- or \pref{eq:RGE-Re} -- is that the value for $C_-/C_+$ --- and so also for all physical observables so obtained  --- actually does not depend on the specific value of $\epscut$ at all, but only on the trajectory $\hlambda(\epscut)$ defined by solving \pref{eq:RGE-Re} given the initial condition $\hat \hlambda_0 = \hlambda(\epscut_0)$.

Since observables depend only on $\epscut$ and $\hlambda(\epscut)$ in an RG-invariant way, it is useful to identify invariant labels for the RG trajectories, since these naturally arise when calculating observables. To this end notice that all of the trajectories in Figure \ref{fig:phase-portraits} cross the imaginary axis at least once, so a convenient choice labels each RG-trajectory using two quantities: 
\begin{enumerate}
\item The scale $\epscut_\star$ where Re$\,\hlambda(\epscut_\star)$ vanishes, and 
\item The imaginary value of $\hlambda(\epscut_\star):=\iu \ystar$ where the crossing occurs. 
\end{enumerate}
The pair $(\ystar,\epscut_\star)$ then provides a convenient RG-invariant parameterization of any flow. When the flow crosses the imaginary axis more than once (as happens only when $\zeta = \iu \xi$ is imaginary) we label the trajectory using the larger of the absolute values for $\ystar$. (Equivalently, we take the crossing for which $\exd \hlambda_\ssR/\exd \epscut > 0$.) This definition ensures $|\ystar| >\xi$, and reduces in the case $\ystar \to \pm\infty$ to the definition of $\epscut_\star$ used in \cite{Burgess2016}. 

In particular, using these definitions in \pref{eq:boundary-ratio} directly gives $C_-/C_+$ in terms of $\epscut_\star$ and $\ystar$: 
\begin{equation}
  \frac{C_-}{C_+}  =  R_\star \; 
   \qty(2k\epscut_\star)^\zeta \e^{\iu \zeta\pi/2}
\qquad \hbox{with} \qquad  R_\star:= R(\epscut_\star)=\frac{\zeta-\iu \ystar}{\zeta+\iu \ystar}
  \label{eq:R-def-re} \,,
\end{equation}
so once observables are expressed in terms of $C_-/C_+$, eq.~\pref{eq:R-def-re} gives them in terms of $\epscut_\star$ and $\ystar$. Notice that the quantity $R_\star$ defined in this expression is either a pure phase when $\zeta$ is real, or is real when $\zeta$ is imaginary. 

For later purposes we also remark that the special case of Hermitian sources considered in \cite{Burgess2016} corresponds to the limits $\ystar=0$ and $\ystar\to\pm \infty$, with these two choices respectively corresponding to the two types of flow found in \cite{Burgess2016}, (distinguished by the relative size of $|\hlambda|$ and $|\zeta|$ in the flow). The invariant $R_\star$ reduces to $R_\star=1$ and $R_\star=-1$ for these two Hermitian classes of flow.\footnote{The sign $-R_\star$ agrees with the RG-invariant sign denoted $y$ in \cite{Burgess2016}.}

\subsubsection*{Fixed points \label{sec:fixed-points}}

Eq.~\pref{eq:RGE-Re} makes clear that there are two fixed points, $\hlambda=\pm \zeta$, for which the coupling $\hlambda$ does not evolve. Given the definitions \pref{prelambdadef} and \pref{lambdadef}, this implies that 
\be
  h(\epscut) \propto \frac{\epscut^{d-2}}{m} 
\ee
at these fixed points, and so in particular $h \propto 1/m$ is $\epscut$-independent when $d=2$, and $h \propto \epscut/m$ when $d =3$. 

Furthermore, eq.~\pref{eq:boundary-ratio} shows the fixed point at $\hlambda = \zeta$ corresponds to setting $C_- = 0$ and so $\psi \propto \psi_+ \propto r^{s_+}$, while the one at $\hlambda = -\zeta$ corresponds to taking $C_+ = 0$ and so $\psi \propto \psi_- \propto r^{s_-}$. 

Finally, notice that $h = 0$ corresponds to $\hlambda = d-2$ and so this is only a fixed point (i.e.\ agrees with $\pm\zeta$) when $g = 0$. Consequently the choice of {\em no} coupling ($h = 0$) is only possible for all $\epscut$ when $g = 0$. More generally, if $h(\epscut_0) = 0$ for some specific $\epscut_0$ then if $g$ is nonzero $h$ necessarily flows away from zero, showing that nonzero $g$ generically precludes the vanishing of $h$ for all scales. Thus, for a generic (i.e.\ $\hlambda\neq \pm \zeta$) boundary condition, imposed at a generic radius $\epscut$, a delta-function is \emph{obligatory}.

The character of the flows when away from the fixed point depend crucially on whether the parameter $\zeta$ is real or imaginary. This in turn depends on $g$, with $\zeta$ real when $g \le g_c$ and $\zeta$ imaginary when $g > g_c$,  where
\be
 g_c := \frac{1}{2m} \left[\varpi_d(\ell) + \frac{(d-2)^2}4\right] \to \left\{ \begin{array}{ll}
  \ell^2/(2m) &\hbox{if $d = 2$}  \\ (\ell + \frac12)^2/(2m)  & \hbox{if $d=3$}  \end{array}  \right. \,.
\ee
Notice in particular that when $d=2$ any positive $g$ satisfies $g > g_c$ for $\ell = 0$. The physical consequence of imaginary $\zeta$, or equivalently $g>g_c$, is that the solutions $\psi_+$ and $\psi_-$ can be viewed as in-falling and out-going waves in logarithmic coordinates (i.e.\ $\psi_\pm=\exp\qty[-\iu (\omega t \pm \frac{\xi}{2} \ln r)]$); this is the quantum manifestation of fall to the centre. 

\subsubsection*{When $\zeta$ is real (sub-critical) \label{sec:disc-subsuper}}

When $\zeta$ is real then all flows begin and end at one of the fixed points, as is clearly seen in the left panel of Figure \ref{fig:phase-portraits}. As $\epscut$ increases the flow is from the UV fixed point at $\hlambda = - \zeta$ to the IR fixed point at $\hlambda = + \zeta$. 

As argued above, from the point of view of the mode functions $\psi(r)$ this corresponds to a crossover from behaviour dominated by $\psi_-(r)$ to that dominated by $\psi_+(r)$. Since for real nonzero $\zeta$ we have $s_+ > s_-$ it follows that $\psi_-(r)$ always dominates for sufficiently small $r$ while $\psi_+(r)$ wins at large-enough $r$. For any particular solution $\psi = C_+ \psi_+ + C_- \psi_-$ the radius for which this crossover happens depends on $C_-/C_+$ (with large values of $C_{-}/C_{+}$ leading to a crossover at larger radii), and it is this crossover that the RG evolution describes.  

If restricted to real $\hlambda$ the two categories of flow found in \cite{Burgess2016} correspond to those that either pass through $\hlambda = 0$ or $ \hlambda = \infty$ when passing between the two fixed points. In this case flows are characterized by only one RG-invariant quantity, $\epscut_\star$, and this can only be much larger than the initial condition, $\epscut_\star \gg \epscut_0$, if the initial condition is chosen very close to the UV fixed point: $\hlambda_0 \simeq - \zeta$. 

For real $\hlambda$ the value $g_c$ is a critical coupling in the sense that as $g$ rises above $g_c$ the two fixed points in the flow of $\hlambda$ coalesce and move into the complex plane as $\zeta$ becomes imaginary, and are consequently inaccessible. We follow the literature ({\em e.g.}~\cite{Moroz2009}) and call the range of $g > g_c$ for which $\zeta^2 < 0$ super-critical while the range $g < g_c$ for which $\zeta^2 > 0$ is sub-critical. The flow equation \pref{eq:RGE-Re} is the poster-child for this kind of transition through which a system loses the existence of a pair of scale-invariant fixed points as a parameter is varied \cite{Kaplan2009}. 

\subsubsection*{When $\zeta$ is imaginary (super-critical)}

The flow topology differs dramatically when $\zeta = \iu \xi$ is imaginary since in this case a generic flow does not have any fixed point. In this case the flows are log-periodic cycles in $\epscut$ that repeat themselves whenever the combination $\ln(\epscut/\epscut_0)$ passes through an integer multiple of $2\pi/\xi$. 

Although $\ystar$ is fixed for a given trajectory, there is an infinite set of values $\epscut_\star$ for which $\hlambda(\epscut_\star) = \iu\ystar$. In particular, for any trajectory it is always possible to find multiple values of $\epscut_\star$ that can be arbitrarily large compared with any particular microscopic scale, like $\epscut_0$.

Once combined with \pref{eq:R-def-re} the existence of multiple $\epscut_\star$'s for fixed $\ystar$ shows that physical quantities will often repeat themselves for different values of $k$.  This limit-cycle behaviour also appears in the multiplicity of Efimov states \cite{Efimov1970}, which when treated using hyperspherical coordinates makes a direct connection with the inverse square potential (see e.g. \cite{BraatenHammer2004}).  

\subsection{The perfect emitter/absorber \label{sec:perfect-abs}}

For imaginary $\zeta$ the fixed points are isolated on the imaginary axis of the complex $\hlambda$ plane, encircled but never reached by other RG trajectories. We now argue that these fixed points correspond to the cases where the source is either a perfect absorber (when Im $\hlambda < 0$ at the critical point) or perfect emitter (when the critical point satisfies Im $\hlambda > 0$). 

This connection to perfect absorption/emission can be understood by realizing that the fixed points correspond to the choices where either $C_-$ or $C_+$ vanish, and so the radial mode functions behave for small $r$ like $\psi \propto \psi_\pm  \propto r^{s_\pm}$. But for $\zeta$ imaginary this behaviour is oscillatory, and periodic in $\log r$, since $\zeta = \iu\xi$ implies
\be \label{perfectabsdef}
  \left( \frac{r}{r_0} \right)^{s_\pm} = \left( \frac{r}{r_0} \right)^{(d-2)/2} \; \exp\left[ \pm\frac{\iu\xi}{2} \; \ln\left( \frac{r}{r_0} \right) \right] \,,
\ee
where we have used the definition of $s_\pm$ given in \pref{spmdef}. This, combined with the time-evolution $e^{-\iu Et}$, shows these solutions can be regarded as in-falling and out-going waves in logarithmic coordinates \cite{Vogt1954,Alliluev1972}. Thus $C_+=0$ (corresponding to $\hlambda = -\iu\xi$) corresponds to the choice of only in-falling waves, a boundary condition that has been used when considering fall to the center \cite{Vogt1954, Alliluev1972}, in-fall of a wave at a black hole horizon \cite{Iqbal2008,Li2012}, and the simulation of wave equations on finite computational domains \cite{Engquis1977,Gander2005}. 

The fact that these boundary conditions are fixed points means that the conditions of perfect absorption or emission both remain completely unchanged as one varies the precise position, $r = \epscut$, at which they are applied.

\section{Scattering, bound states, and RG invariants}
\label{sec:RG-obs}

The purpose of this section is to establish an explicit connection between observables and the RG-invariant parameters $\epscut_\star$ and $\ystar$ (or equivalently $R_\star$ as defined in \pref{eq:R-def-re}), for later use when connecting theory to experimental applications. This connection is discussed in many places in the literature for real $\hlambda$ \cite{Gupta:1993id, Beane2000, Coon2002, Bawin2003, Camblong2003, Braaten2004, Hammer2006, Bouaziz2014, Burgess2016}, for which the scattering length, $a_s$, turns out to be given in terms of the two RG-invariants, $\epscut_\star$ and $R_\star = \pm 1 = -  \hbox{sgn} (|\hlambda| - |\zeta|)$, by the relation \cite{Burgess2016}
\be \label{realasepsstar}
 a_s = - R_\star \epscut_\star \qquad (\hbox{for real $\hlambda$}) \,. 
\ee

A relation like \pref{realasepsstar} (for real $\hlambda$) makes systems where $\epsilon_\star$ is large relative to the source (i.e.\ $\epscut_\star\gg r_p$), particularly interesting. This is because such systems have anomalously large scattering lengths compared to their underlying dimensions. Large $\epscut_\star$ occurs for real RG trajectories whose microscopic initial condition, $\hlambda(\epscut_0)$, lies very close to the UV fixed point, $\hlambda(\epscut_0) \simeq - \zeta$. Due to the slow flow in the vicinity of the fixed point (as seen, for example, in Figure \ref{fig:phase-portraits}) this type of initial condition makes $\epscut_\star$ exponentially larger than $\epscut_0$ \cite{Burgess2016}, and the RG effectively re-sums the effects of what is becoming a large coupling. This is the first-quantized version of what is also found in second-quantized formulations \cite{Weinberg:1990rz, Kaplan:1996xu, Mehen:1998zz, Braaten2004} for systems with large scattering lengths.

We next generalize these observations to non-self-adjoint actions, so they may be applied to the physics of absorptive point-like sources (like the charged wire considered in detail below). We find that absorptive sources display similar behaviour, with the amount of non-unitary scattering being characterized by both $\epscut_\star$ and the value of $\ystar$.

\subsection{Scattering \label{sec:scattering}}

We start in this section by formulating scattering observables when the physics is partially absorptive.

\subsubsection{Elastic and absorptive scattering cross sections  
\label{ssec:4-ela-abs}}

Appendix \ref{app:abs-sc} (and references therein) review the main results and conventions for absorptive scattering. The main quantity to be computed is the $S$-matrix element for each partial wave, $S_\ell = \e^{2\iu\gamma_\ell}$, where for absorptive scattering the phase shift is complex: $\gamma_\ell=\delta_\ell + \iu \eta_\ell$. The limit of unitary scattering corresponds to $\eta_\ell \to 0$. 

The elastic cross section for each partial wave is defined in the usual way, and is given in terms of $\delta_\ell$ and $\eta_\ell$ by
\begin{equation} \label{eq:sigma-el3}
  \sigma^{(\ell)}_{\textrm{el}}
  =\frac{\pi(2\ell +1)}{k^2}
  \Bigl[1+ \e^{-4\eta_\ell} -2 \e^{-2\eta_\ell} \cos 2\delta_\ell \Bigr]
  \qquad \hbox{if $d=3$}
\end{equation}
and
\begin{equation}    \label{eq:sigma-el2}
  \sigma^{(\ell)}_{\textrm{el}}
  =\frac{1}{k} \Bigl[1+ \e^{-4\eta_\ell} - 2 \e^{-2\eta_\ell} \cos 2\delta_\ell\Bigr]
  \qquad\qquad\qquad \hbox{if $d=2$} \,.
\end{equation}

The absorptive cross sections for each partial wave are similarly given by (see Appendix \ref{app:abs-sc} or \cite{Adhikari1986} for more details)
\begin{equation}
  \sigma^{(\ell)}_{\textrm{abs}} 
  =\frac{\pi(2\ell +1)}{k^2}
  \Bigl( 1- \e^{-4\eta_\ell} \Bigr) \qquad \hbox{if $d=3$}
   \label{eq:sigma-abs3}
\end{equation}
and
\begin{equation}
   \sigma^{(\ell)}_{\textrm{abs}} 
   =\frac{1}{k} \Bigl( 1- \e^{-4\eta_\ell} \Bigr) \qquad\qquad~ \hbox{if $d=2$}\,.
   \label{eq:sigma-abs2}
\end{equation}
Recall that in these formulae $\ell\in\mathbb{Z}$ when $d = 2$ while for $d=3$ we instead have $\ell\in\mathbb{N}$. These become standard results in the limit of unitary scattering, $\eta_\ell \to 0$, for which in particular $\sigma^{(\ell)}_{\rm abs} \to 0$. 

For later purposes we call the limit $\eta_\ell\rightarrow\infty$ the case of `maximal' or `total' absorption. In this limit the above formulae show that $\sigma^{(\ell)}_{\rm abs} = \sigma^{(\ell)}_{\rm el}$,  which is a manifestation of the fact that all ingoing spherical waves are absorbed. This can be understood by noting that an incident plane wave is composed of equal parts outgoing and ingoing spherical waves, and so half of the incident probability flux is absorbed and the other half is scattered. As becomes clear below, maximal absorption is {\em not} the same as the perfect-absorber criterion defined near eq.~\pref{perfectabsdef}.

\subsubsection{Relation to RG invariants}

It remains to obtain the two quantities $\e^{-4\eta_\ell}$ and $\e^{-2\eta_\ell}\cos 2\delta_\ell$ from the RG invariants $\epscut_\star$ and $\ystar$. To avoid clutter from here on we suppress dependence on the ubiquitous subscript $\ell$ appearing in \cref{eq:sigma-el3,eq:sigma-el2,eq:sigma-abs3,eq:sigma-abs2} in these expressions.

Repeating the steps of \cite{Burgess2016} relating $C_+/C_-$ to the unitary phase shift, and specializing to the cases $d=2$ and $d=3$, leads to
\begin{align}
  \e^{2 \iu \gamma}= \left[ \frac{1+\cA\,\e^{\iu\pi\zeta/2}}
  {1+\cA\,\e^{-\iu\pi\zeta/2}} \right] \e^{\iu\pi(2\ell+1-\zeta)/2}
   \qquad \hbox{if $d=3$} \label{eq:phase-shift-3}\\
   \hbox{and} \quad \e^{2\iu\gamma}= \left[ \frac{1+\cA\,\e^{\iu\pi\zeta/2}}
  {1+\cA\,\e^{-\iu\pi\zeta/2}} \right] \e^{\iu\pi(2\ell-\zeta)/2 }
    \qquad~~~ \hbox{if $d=2$} \label{eq:phase-shift-2}
\end{align}
where $\cA$ is defined via \cite{Burgess2016} 
\begin{equation}
 \cA 
 := \frac{\e^{-\iu\pi\zeta/2}}{2^{2\zeta}} \left( \frac{C_-}{C_+} \right) 
  \frac{ \Gamma\qty(1-\tfrac{1}{2}\zeta) }{ \Gamma\qty(1+\tfrac{1}{2}\zeta) }
  = R_\star \qty(\frac{k\epscut_\star}{2})^\zeta
  \frac{\Gamma\qty(1-\tfrac{1}{2}\zeta)}{\Gamma\qty(1+\tfrac{1}{2}\zeta)} \,, 
  \label{eq:4-ela-abs-10}
\end{equation}
and we use, as before, the definition $R_\star := (\zeta - \iu\ystar)/(\zeta + \iu\ystar)$ given in \pref{eq:R-def-re}. Combined with \pref{eq:boundary-ratio} we obtain the second equality in \pref{eq:4-ela-abs-10}.

Eqs.\ \pref{eq:phase-shift-3} and \pref{eq:phase-shift-2} are particularly simple in the special case of real $\zeta$ and evaluated at the IR fixed point, for which $C_-=0$ and so ${\cal A} = 0$.  In this case eqs.\ \pref{eq:phase-shift-3} and \pref{eq:phase-shift-2} imply the phase shift $\gamma= \delta$ is real, as expected given that the imaginary part of $\hlambda$ always vanishes at the UV fixed point. Using the convenient feature that the inverse square potential has the same functional form as the centrifugal barrier, we define the effective angular momentum by 
\be \label{deltaelldef}
 \zeta=2\ell_\text{eff}+1 \quad \hbox{(for $d=3$)} \qquad \hbox{and} \qquad \zeta=2\ell_\text{eff} \quad \hbox{(for $d=2$)}.
\ee
This allows us to see that\footnote{This is as would be expected by comparing the asymptotic form of a free spherical wave (e.g.  $H_\ell^{(1)}\sim r^{-1/2}\e^{\iu(kr-\ell\pi/2)}$ in $d=2$) with one whose effective angular momentum is modified by the inverse square potential (e.g.  $H_{\ell_\text{eff}}^{(1)}\sim r^{-1/2}\e^{\iu(kr-\ell_{\text{\eff}}\pi/2)}$ in $d=2$).}  $\delta= (\ell-\ell_\text{eff})\pi/2$, in agreement with \cite{Alliluev1972}.

We next examine the cases where $\zeta$ is real or imaginary separately. In the super-critical case we present an explicit comparison of the parametric requirements for total absorption {\em vs.} a perfect-absorber. 

\subsubsection*{When $\zeta$ is real (sub-critical)}

We begin by expressing the modulus and phase of $\cA = \modA \, \e^{\iu\argA}$ in terms of the RG invariant quantities $\epscut_\star$ and $\ystar$, keeping in mind that the reality of $\zeta$ means that the quantity $R_\star$ is a pure phase. We therefore find
\begin{align}
  \modA&=\qty(\frac{k\epscut_\star}{2})^\zeta
  \abs{\frac{\Gamma\qty(1-\tfrac{1}{2}\zeta)}
    {\Gamma\qty(1+\tfrac{1}{2}\zeta)}}\\
  \argA&= \sgn(-\ystar)\arccos\left( \frac{\zeta^2-\ystar^2}{\zeta^2+\ystar^2} \right)
  + \frac{\pi}{2}\left\{ \textrm{sgn}\left[
  \frac{\Gamma\qty(1-\tfrac{1}{2}\zeta)}{\Gamma\qty(1+\tfrac{1}{2}\zeta)} \right]
    -1 \right\} \,,
  \label{eq:rho-and-phi-sub}
\end{align}
where the $\text{sgn}$ function arises as a consequence of the condition that $\modA>0$. In this case $\epscut_\star$ appears only in $\modA$ while $\ystar$ appears only in $\argA$. 

We can use \pref{eq:phase-shift-3} and \pref{eq:phase-shift-2} to express the complex scattering phase in terms of $\modA$ an $\argA$, leading to the following expressions 
\begin{align}
  \e^{-4\eta}&=\frac{1+2\modA\cos\qty(\argA+\frac{\pi\zeta}{2})+\modA^2}
  {1+2\modA\cos\qty(\argA-\frac{\pi\zeta}{2})+\modA^2}
  \label{eq:4-ela-abs-12.1} \\
  \e^{-2\eta}\cos 2\delta&=\frac{\cos(\Delta \pi)
    +2\modA\cos\argA\cos\qty(\Delta \pi+\frac{\pi\zeta}{2})
    +\modA^2\cos(\Delta \pi+\pi\zeta)}
      {1+2\modA\cos\qty(\argA-\frac{\zeta\pi}{2})+\modA^2 }
  \label{eq:4-ela-abs-12.2}
\end{align}
where $\Delta =\ell-\ell_\textrm{eff}$, with $\ell_{\rm eff}(\zeta)$ defined in \pref{deltaelldef}, is introduced in order to make this expression the same for both $d=2$ and $d=3$.

The absorptive cross section for each partial wave is then given in terms
of $\zeta$, $\modA$ and $\argA$ by \pref{eq:sigma-abs3} and \pref{eq:sigma-abs2} with
\begin{equation}
1-\e^{-4\eta}=\frac{4\modA\sin\argA\sin\frac{\pi\zeta}{2}}
  {1+2\modA\cos\qty(\argA-\frac{\pi\zeta}{2})+\modA^2} \,,
  \label{eq:4-ela-abs-13.1}
\end{equation}
from which it can be seen that the absorptive cross section vanishes in both the limits $\modA\rightarrow 0$ and $\modA\rightarrow \infty$ with the other parameters fixed. It also vanishes if $\argA = 0$ or $\pm\pi$, as is the case when $\ystar=0$ or $\ystar=\pm\infty$ respectively (the choices corresponding to the purely real $\hlambda$ considered in \cite{Burgess2016}). 
 
Naively, eq.~\pref{eq:4-ela-abs-13.1} seems also to imply that the
absorptive cross section vanishes (i.e.\ $\eta \to 0$) when $\zeta$ passes through zero,
and a similar conclusion seems also to follow from \pref{eq:phase-shift-3} and \pref{eq:phase-shift-2}. This conclusion turns out to be incorrect, however, due to the hidden dependence of other variables on $\zeta$. The subtlety of this limit can be seen from \pref{eq:4-ela-abs-10}, which shows $\cA \to -1$ in the limit $\zeta \to 0$, provided $\ystar=\ystar^{(0)}+\order{\zeta}$,  making the expressions \pref{eq:phase-shift-3} and \pref{eq:phase-shift-2}  indeterminate. This limit is tricky because our choice of mode functions $\psi_\pm \propto r^{s_\pm}$ degenerates to become linearly dependent 
when $\zeta \to 0$.  

\begin{figure}
  \centering
  \begin{subfloat}[] [$\zeta=2.15$]%
    {\includegraphics[clip,width=0.5\linewidth]{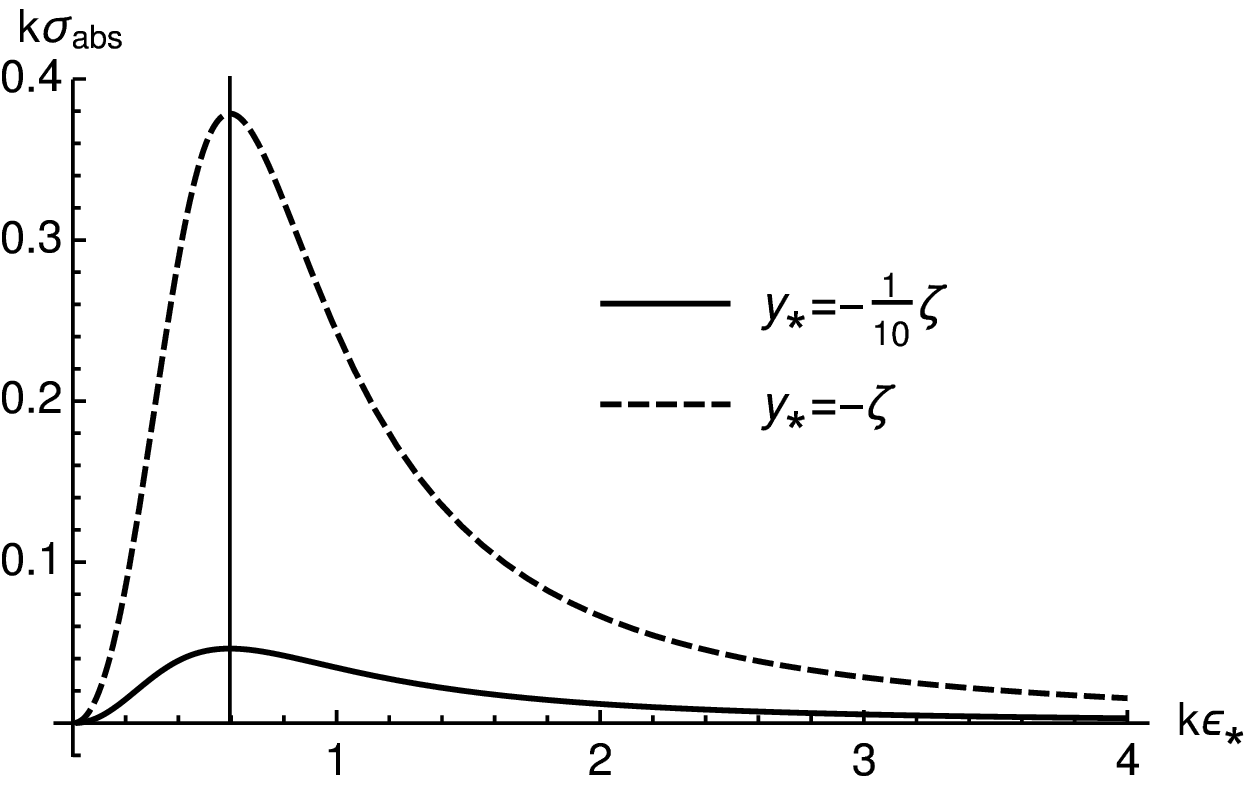} 
      \label{fig:sub-k}} %
  \end{subfloat}%
  \begin{subfloat}[][$\zeta=\iu\xi$ with $\xi=2$]%
   {    \includegraphics[clip,width=0.5\linewidth]{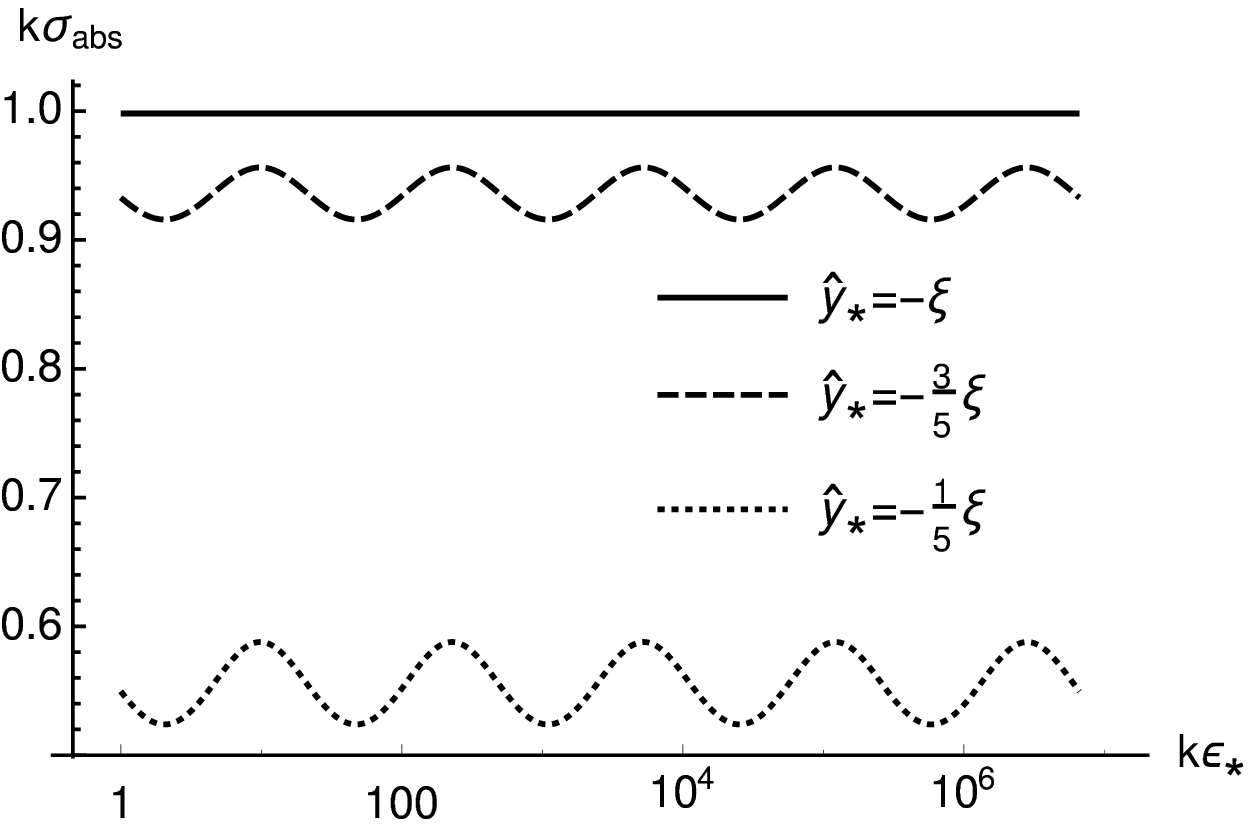}    
     \label{fig:super-k} }%
 \end{subfloat} 
 \caption{Illustration of the breaking of scale-invariance for $\zeta$ real (i.e.\ $g<g_c$) \protect\subref{fig:sub-k} and $\zeta=\iu \xi$ imaginary (i.e.\ $g>g_c$)\protect\subref{fig:super-k}. The dimensionless combination $k\sigma_\text{abs}$ [see \pref{eq:sigma-abs2}] is plotted as a function of $\zeta$ using \pref{eq:4-ela-abs-13.1} and \pref{absxi}. For $\zeta$ real we see that the characteristic value of $k$ which dominates the absorption is set by the RG-invariant scale $k_\text{res}\approx 0.5/\epscut_\star$. The resonant momentum calculated in \pref{eq:kres} is shown as a vertical line. Provided the remainder of $\zeta/2$ is bigger than about $0.1$ the  width and shape of the peak is determined exclusively by $\epscut_\star$ and $\zeta$, while its height is determined by $\ystar$. For $\zeta=\iu \xi$ imaginary the absorptive cross section displays a \emph{discrete} scale invariance, being periodic in $\log k$. This provides a particularly simple experimental signature of a quantum anomaly. Note that in the classical limit $k\sigma_\text{abs}$ has no scale dependence (see \cref{app:cl-abs}). \label{fig:super-vs-sub-k}}
\end{figure}

The behaviour of the absorptive cross section as a function of $k\epscut_\star$ is illustrated in the left panel of Figure \ref{fig:super-vs-sub-k}. This shows how the biggest effects arise when $k \epscut_\star$ is of order unity, which is only possible (again) within the EFT regime when $\epscut_\star$ is much larger than the underlying microscopic size of the source.

\subsubsection*{When $\zeta$ is imaginary (super-critical)}

When $\zeta = \iu \xi$ is imaginary we have
\begin{equation}
 \cA=  \qty(\frac{k\epscut_\star}{2})^{\iu\xi}
 \left( \frac{ \xi-\ystar }{  \xi+\ystar} \right)
  \frac{\Gamma\qty(1-\tfrac{\iu}{2}\xi)}{\Gamma\qty(1+\tfrac{\iu}{2}\xi)}
  \label{eq:4-ela-abs-13} \,,
\end{equation}
and so the modulus and phase of $\cA= \modA \,\e^{\iu\argA}$ are given by
\begin{equation} \label{modAxiystar}
  \modA=\frac{\ystar-\xi}{\ystar+\xi}
\end{equation}
and
\begin{equation}\label{eq:argA-supercrit}
  \begin{split}
    \argA&=\pi+\xi\ln\left( \frac{k\epscut_\star}{2}\right)+
    2~\textrm{arg}~\Gamma\qty(1-\tfrac{1}{2}\iu\xi)\\
    &=\pi+\xi\ln\left( \frac{k\epscut_\star}{2}\right)+\gamma_\ssE+\sum_{s=0}^{\infty}\left[
    \frac{\xi}{1+s}-2\arctan\qty(\frac{\tfrac{1}{2}\xi}{1+s})\right] \\
    &\simeq \pi+ \gamma_\ssE+\xi\ln\left( \frac{k\epscut_\star}{2}\right)+ \frac{\xi^2}{6} + \mathcal{O}(\xi^3) 
    ~~~~\text{(for small}~\xi)\,,
  \end{split}
\end{equation}
where $\gamma_\ssE=0.5772$ is the Euler-Mascheroni constant. In this case it is $\ystar$ that determines the modulus of $\cA$ while $\epscut_\star$ controls its phase. Notice that $\modA \ge 0$ because of our convention that $|\ystar| \ge \xi$.

In this case the complex scattering phase is given by $\gamma = \delta + \iu \eta$ with
\begin{align}
  \e^{-4\eta}&=\frac{\e^{\xi\pi/2}~~+~2\modA\cos\argA~+\e^{-\xi\pi/2}\modA^2}
    {\e^{-\xi\pi/2}+~2\modA\cos\argA~+~\e^{\xi\pi/2}\modA^2}
  \label{eq:4-ela-abs-14.1}\\
  \e^{-2\eta}\cos{2\delta}&=
  \frac{(1+\modA^2)\cos\frac{\zeta_0\pi}{2}+\modA \,\e^{\xi\pi/2}\cos(\frac{\zeta_0\pi}{2}+\argA)
    +\modA \,e^{-\xi\pi/2}\cos(\frac{\zeta_0\pi}{2}-\argA) }
       {(\e^{-\xi\pi/4}+\modA \e^{\xi\pi/4})^2}
       \label{eq:4-ela-abs-14.2}
\end{align}
where $\zeta_0=\sqrt{(d-2)^2+4\varpi_d(\ell)}$ is the value of $\zeta$ when $g = 0$, which in the familiar cases of $d=2$ and $d=3$ is given by $\zeta_0=2\ell$ and $\zeta_0=2\ell+1$ respectively. The absorptive cross section in this case is proportional to
\be
   1-\e^{-4\eta} = \frac{2(\modA^2 - 1) \sinh({\xi\pi/2})}
    {\e^{-\xi\pi/2}+~2\modA\cos\argA~+~e^{\xi\pi/2}\modA^2} \,,
  \label{absxi}
\ee
which vanishes for $\modA= \pm 1$ (as again corresponds to the Hermitian case, $\ystar = 0$ or $\ystar \to \pm \infty$, as is seen from \pref{modAxiystar}).

These expressions also show the difference between total absorption (defined by $\eta \to \infty$, such as is obtained if $\cos \argA = -1$ and $\modA = \e^{\xi\pi/2}$) and a perfect absorber (defined by the fixed point, $\ystar = - \xi$, for which $\modA \to \infty$). In the case of total absorption we have 
\be
   \sigma^{(\ell)}_{\rm el} = \sigma^{(\ell)}_{\rm abs} = f_d(k) \qquad \hbox{(total absorber)}\,,
\ee
where $f_3(k) = \pi (2\ell + 1)/k^2$ for $d = 3$ and $f_2(k) = 1/k$ for $d=2$.  

A perfect absorber, on the other hand, predicts $\eta = \xi\pi/4$ and so $\delta = \zeta_0 \pi /4$, leading to
\begin{equation}  
  \sigma^{(\ell)}_{\textrm{el}}  = f_d(k)
  \left[1+ \e^{-\xi\pi} -2\e^{-\xi\pi/2}\cos \frac{\zeta_0\pi }{2} \right]
  \qquad \hbox{(perfect absorber)}
\end{equation}
and
\begin{equation} \label{perfectabsform}
  \sigma^{(\ell)}_{\textrm{abs}} 
  =f_d(k) \Bigl( 1- \e^{-\xi\pi} \Bigr) \qquad\qquad\qquad\qquad \hbox{(perfect absorber)} \,.
\end{equation}
As is shown in Appendix \ref{app:cl-abs} this perfect-absorber limit goes over to standard formulae for perfect classical absorption \cite{Alliluev1972}
\be \label{classlimperfabs}
  \sigma_\text{abs} \approx \frac{\pi \ell_c^2}{k^2} = \pi b_c^2 \quad \hbox{(if $d=3$)} \qquad \hbox{and} \qquad
  \sigma_\text{abs} \approx \frac{2 \ell_c}{k} =2 b_c \quad \hbox{(if $d=2$)} \,,
\ee
in the limit where the critical angular momentum --- defined by the value for which $\zeta(\ell_c)$ is closest to zero --- is large: $\ell_c \gg 1$. Here $b_c = \ell_c /k$ is the classical impact parameter that leads to contact with the wire.

More generally, in the super-critical regime, the absorptive cross section has an oscillatory structure when viewed as a function of $\log k$, as can be seen from the right panel of Figure \ref{fig:super-vs-sub-k}. This kind of periodic behaviour is a consequence of the inverse square potential's breaking of a continuous scale-invariance to a discrete subgroup \cite{Hammer2006,Kaplan2009,Moroz2009}.
 
The residual discrete symmetry is most easily understood by returning to the phase portraits of Figure \ref{fig:phase-portraits} where distinctive limit cycle behaviour can be seen. A consequence of the closed orbits is that for a fixed  value of $\ystar$ there are a countably infinite number of energy scales $\{\epscut_\star^{(n)}\}$ satisfying  $\epscut_\star^{(n)}/\epscut_\star^{(n+1)}=\e^{-2\pi/\xi}$ [see \pref{eq:argA-supercrit}]. In particular, there is no need to choose special initial conditions at microscopic scales, $\epscut_0$, in order to ensure the existence of macroscopically large $\epscut_\star$. The experimental signature of this fact is the behaviour seen in the absorption rate for a given partial wave, which is a log-periodic function of the incident particle's energy with periodicity $\log k\rightarrow \log k + 2\pi/\xi$. 

\subsection{Bound states}

Bound states --- defined as negative energy solutions to the time-independent Schr\"odinger equation that are normalizable at infinity --- make up a second important class of observables. For non-self-adjoint systems such states can have complex energy eigenvalues, corresponding to the bound state's decay (or growth) due to its repeated access to the relevant non-unitary physics at the source. 

Writing the complex energy eigenvalue as $\mathcal{E}=E-\iu\Gamma/2$, the imaginary part $\Gamma \ge 0$ is to be interpreted as the width (or inverse mean-lifetime) of the bound state. In terms of its modulus and phase, $\cE = \modE \, \e^{\iu\argE}$, for bound states we have $E = \modE \cos \argE < 0$ and $\frac12 \, \Gamma= -\modE \sin \argE \ge 0$ and so $\cE$ resides in the lower-left quadrant: $\pi\le \argE< 3\pi/2$. For such systems it is also convenient to introduce a complex momentum, related to the energy by $\mathcal{E}=\mathcal{K}^2/2m$.

Normalizability at infinity is imposed by demanding that the mode solution $C_+\psi_+ + C_-\psi_-$ decay sufficiently quickly at infinity. Because both $\psi_+(\mathcal{K}r)$ and $\psi_-(\mathcal{K}r)$ grow exponentially when $\abs{\mathcal{K}r}\gg 1$ (for $\text{arg}~\mathcal{K}r\neq\pm \pi/2$) this is only possible for a specific choice for $C_-/C_+$ which must be chosen such that these exponential pieces cancel. This leads to the condition ({\em c.f.}~reference \cite{Burgess2016})
\begin{equation}\label{eq:bs-cond}
  \frac{C_+}{C_-}=\frac{\Gamma\qty(+\tfrac{1}{2}\zeta)}{\Gamma\qty(-\tfrac{1}{2}\zeta)}\,,
\end{equation}
which, when combined with \pref{eq:R-def-re} yields
\begin{equation}\label{ComplexEeq1}
  \frac{\zeta-\iu\ystar}{\zeta+\iu\ystar}=
  \frac{\Gamma\qty(+\tfrac{1}{2}\zeta)}
       {\Gamma\qty(-\tfrac{1}{2}\zeta)}
       \qty(\frac{\mathcal{K}\epscut_\star}{2})^{-\zeta}\e^{\iu \pi\zeta/{2}} \,.
\end{equation}
This expression can be solved for $\cK$ (or $\cE$) as a function of $\ystar$, $\zeta$ and $\epscut_\star$, with result
\begin{equation}\label{ComplexEeq}
  \cE = -\frac2{m\epscut^2_\star}\left[ \left( \frac{\zeta-\iu\ystar}{\zeta+\iu\ystar} \right) 
  \frac{\Gamma\qty(-\tfrac{1}{2}\zeta)}
       {\Gamma\qty(+\tfrac{1}{2}\zeta)} \right]^{-2/\zeta}
         \,.
\end{equation}
Notice that this expression can only be trusted if it lies within the EFT limit, for which $|\cK \epscut_0| \simeq |\cK \rp| \ll 1$, and so, whenever \pref{ComplexEeq1} implies $\cK \epscut_\star \simeq \cO(1)$, we require $\epscut_\star \gg \epscut_0, \rp$.

\subsubsection*{When $\zeta$ is real (sub-critical)}

For $\zeta$ real \pref{ComplexEeq} implies the phase and amplitude, respectively, of the complex bound state energies are 
\begin{align}\label{eq:epsilons-sub}
  \argE&= \sgn(-\ystar)\qty[\frac{1}{\zeta}~ 2\arccos\left( \frac{\zeta^2-\ystar^2}{\ystar^2+\zeta^2} \right)] +\pi \\
 \modE&=\frac{2}{m\epscut^2_\star}
 \left[ \frac{\Gamma\qty(-\tfrac{1}{2}\zeta)}
     {\Gamma\qty(+\tfrac{1}{2}\zeta)} \right]^{-2/\zeta} \,,\label{eq:epsilons-sub2}
\end{align}
and these predictions for $E$ and $\Gamma$ are graphed as a function of $\zeta$ in the left panel of Fig.~\ref{fig:bound-states}.

We see that the decay rate $\Gamma=0$ when either $\ystar = 0$ or $\ystar=\pm\infty$ (as expected for these two distinct unitary limits), and that the single bound state obtained in the latter case is the usual one supported by a $\delta$-function potential. The existence of the single bound state would be hard to understand if one considered the inverse square potential in isolation \cite{Meetz1964}, because the effective radial potential, ($\varpi_d(\ell)-2mg)/r^2$, is typically repulsive for the sub-critical case, $\zeta^2>0$, and so we refer to this feature as an `exotic bound state'. Only more recently has it been appreciated that this state is supported by the $\delta$-function potential \cite{Bouaziz2014, Burgess2016}, rather than the inverse-square potential. 

As discussed in \S\ref{sec:renormalization}, for the predicted bound state of eq.\ \pref{eq:epsilons-sub2} to lie within the regime of validity of our effective treatment we require  $\rp \ll\epscut_0 \ll \epscut_\star$, where $\epscut_\star$ plays the role of $a$ (since the two are in one-to-one correspondence). This ensures that the spatial extent of the bound state is much larger than the compact object that resides at the origin. We may relate  $\epscut_\star$ to $\epscut_0$ by noting that the left-hand side of \pref{eq:boundary-ratio} is independent of $\epscut$, and so by evaluating the right hand side at both $\epscut_0$ and $\epscut_\star$ we obtain the following relationship (\emph{c.f.} reference \cite{Burgess2016}) 
\begin{equation}
   \epscut_\star = \epscut_0 \qty (\frac{\mathfrak{d}_0-2}{R_\star \mathfrak{d}_0} )^{1/\zeta},
   \label{eq:epsstar-eps-rel}
\end{equation}
where $\mathfrak{d}_0:=\hlambda(\epscut_0) + \zeta$ measures the distance between the initial condition $\hlambda(\epscut_0)$ and the  UV fixed point $\hlambda=-\zeta$. There are two ways in which the hierarchy $\rp \ll \epscut_0 \ll \epscut_\star$ arises naturally. First, if the initial conditions are extremely close to the UV fixed point (i.e.\ $\mathfrak{d}_0\ll 1$) then $\epscut_\star/\epscut_0\gg 1$ by virtue of the small denominator in \pref{eq:epsstar-eps-rel}.  Alternatively, for $\mathfrak{d}_0$ small, but not infinitesimal (i.e.\ $\mathfrak{d}_0\lesssim 1$), we may also have $\epscut_\star\gg \epscut_0$ provided that $\zeta \ll 1$, because of the exponent $1/\zeta$ in \pref{eq:epsstar-eps-rel}. This is a particularly interesting observation, because, in the case of the charged wire example, the value of $\zeta$ can be tuned arbitrarily close to zero in the lab. With this application in mind, we expand eqs.\ \pref{eq:epsilons-sub} and \pref{eq:epsilons-sub2} for $\zeta\ll 1$ taking care to choose the appropriate branch for the phase of the complex energy
\begin{align}\label{eq:epsilons-sub-small}
  \argE&\approx \qty[\frac{4}{\abs{\ystar}} + \frac{4}{3 \abs{\ystar}^3} \zeta^2] +\pi \\
 \modE&\approx\frac{2}{m\epscut^2_\star} \e^{-2\gamma_\ssE} \qty(1+\frac{\psi^{(2)}(1)}{12} \zeta^2),\label{eq:epsilons-sub-small-2}
\end{align}
where $\gamma_\ssE\approx 0.5772$ is the Euler-Mascheroni constant, and $\psi^{(2)}(1)\approx -2.404$ is the polygamma function of order two evaluated at one.

\subsubsection*{When $\zeta$ is imaginary (super-critical)}

\begin{figure}
  \centering
  \begin{subfloat}[$\zeta$ real]%
  {  \includegraphics[clip,width=0.45\linewidth]{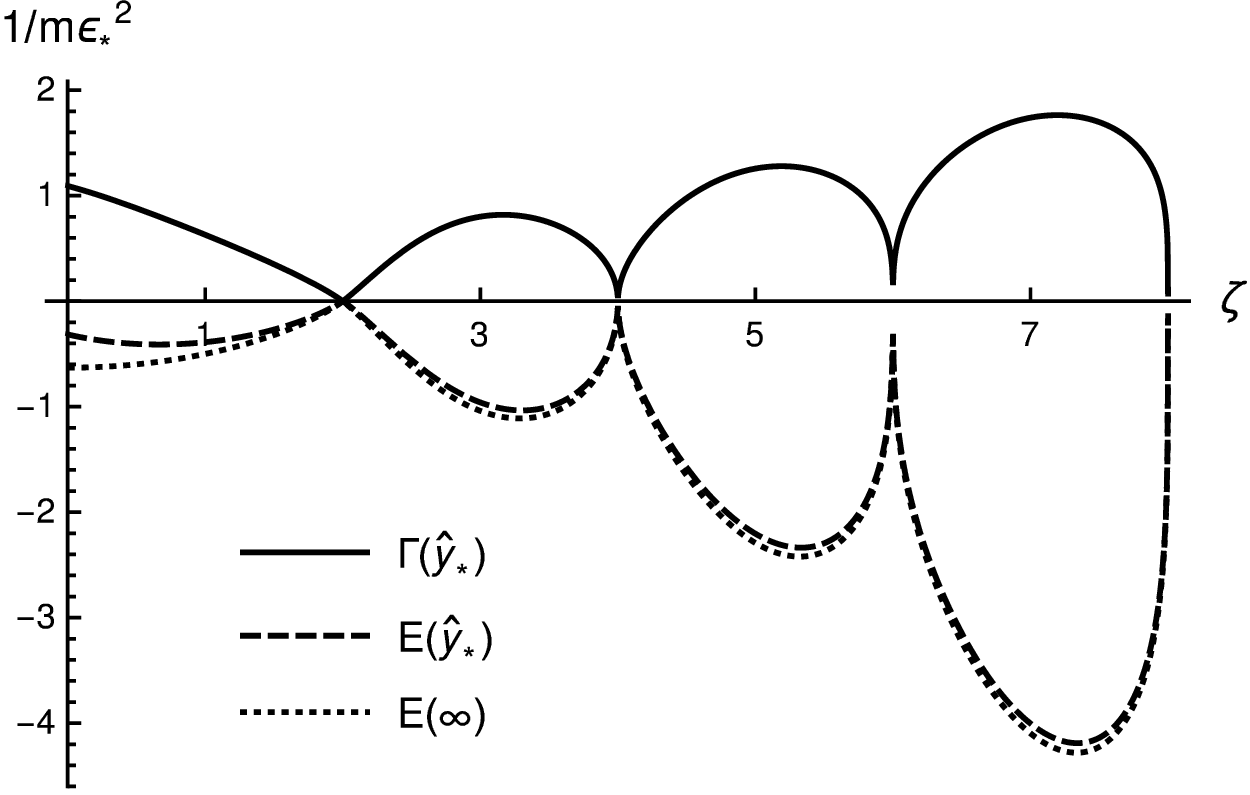}
    \label{fig:exotic-bs}  }
  \end{subfloat}%
 \begin{subfloat}[$\zeta=\iu\xi$ imaginary]%
   {    \includegraphics[clip,width=0.45\linewidth]{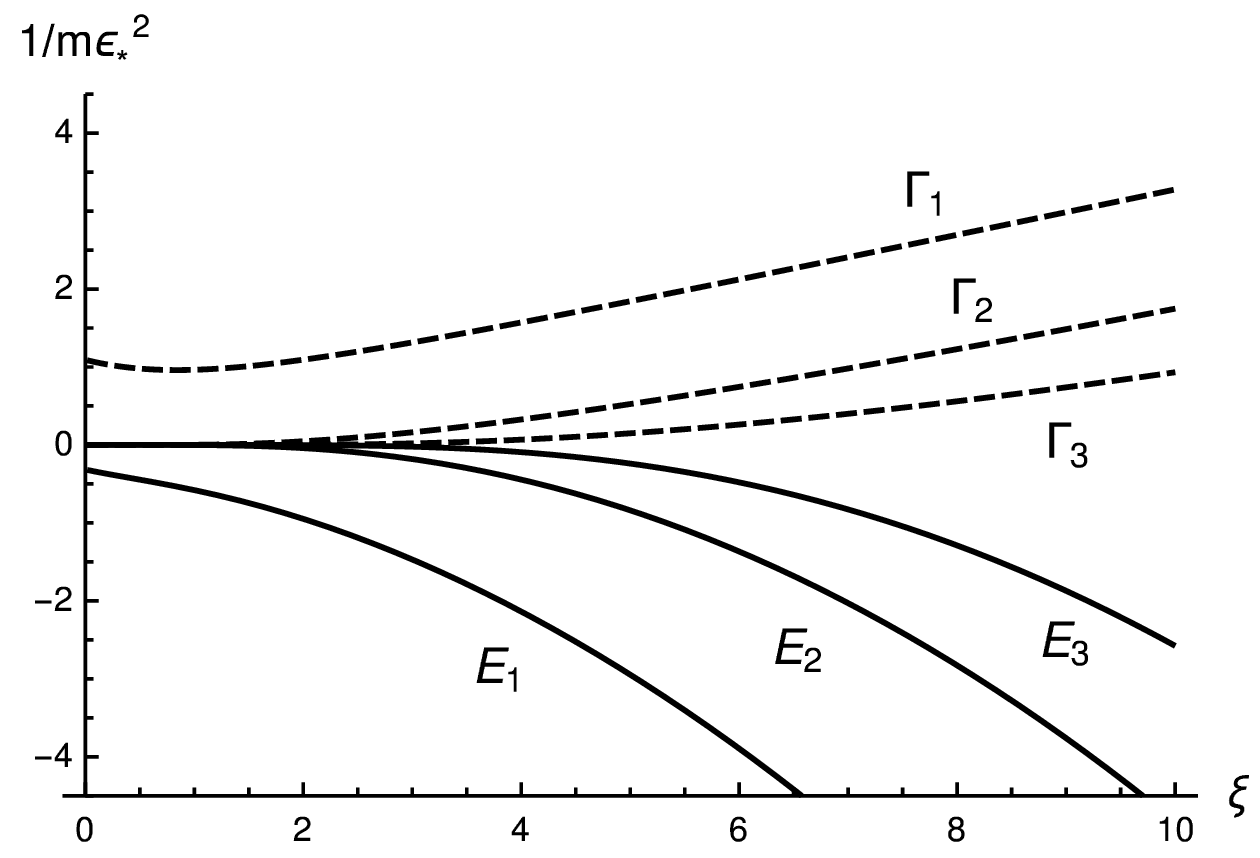} 
    \label{fig:efimov-bs}} %
 \end{subfloat}
 \caption{ Energy $E(\zeta)$ and inverse lifetimes $\Gamma(\zeta)$ of bound states are plotted as a function of $\zeta$. Panel \protect\subref{fig:exotic-bs}: for real $\zeta$ values for $E$ and $\Gamma$ for $\ystar=-12/\pi-2\zeta$ are shown (see \pref{eq:epsilons-sub} and \pref{eq:epsilons-sub2}), and the energy of the unitary (i.e.\ $\ystar=-\infty$) bound state is included for comparison. The zeros in \protect\subref{fig:exotic-bs} are related to the zeros of the Gamma function as can be seen from \pref{eq:epsilons-sub2}. We emphasize that for $\zeta$ real only one bound state exists for each $\ystar$. Panel \protect\subref{fig:efimov-bs}: for imaginary $\zeta=\iu\xi$ the same quantities $E$ and $\Gamma$ (see \pref{eq:arge-im} and \pref{eq:mode-im}) are plotted for $\ystar=-12/\pi-2\xi$. Three different states in the Efimov-like tower are shown, each related to one another via $E_{n+1}=\e^{-4\pi/\xi}E_n$ [since $E\propto 1/\epscut_\star^2$ as can be seen in \pref{eq:mode-im}]. Here there are an infinite tower of bound states for each $\ystar$ in contrast to the case of $\zeta$ real. We note that only one state in the tower can have a nonzero limit as $\xi\rightarrow 0$, and in this case that state is labelled as $E_1$ and $\Gamma_1$, which  are seen to be continuously connected to the corresponding values plotted in  \protect\subref{fig:exotic-bs}. \label{fig:bound-states} }

\end{figure}

Many more bound states can exist when $\zeta =\iu \xi$ is imaginary, since then the limit cycle behaviour implies a countably infinite tower of Efimov-like \cite{Efimov1970,Fonseca1979,Kraemer2006,Braaten2007,Werner2008,Platter2009,Hammer2010,Hammer2011,Macneill2011} narrow bound states satisfying
\begin{align}\label{eq:arge-im}
  \argE&=\frac{2}{\xi}
  \log\qty[\frac{\ystar-\xi}{\ystar+\xi}]+\pi\\
  \modE&= \frac{2}{m\epscut_\star^2}\exp\qty[\frac{2}{\xi}\qty( [2 n
  +1]\pi+\arg \frac{\Gamma(+\tfrac{\iu\xi}{2}) }{\Gamma(-\tfrac{\iu\xi}{2})}  )],\label{eq:mode-im}
\end{align}
where we choose the appropriate branch by specifying a choice for the integer $n(\xi)$ for a given value of $\xi$. The corresponding implications for $E$ and $\Gamma$ are plotted in the right-hand panel of Fig.~\ref{fig:bound-states}.

The integer $n$ here labels the different bound states, whose energies are related to one another by a discrete scale transformation $E \rightarrow \e^{-4\pi/\xi} E$ [since $E\propto 1/\epscut_\star^2$ as shown in \pref{eq:mode-im}] due to the requirement that $\cE$ remain continuous as one moves from one sheet to the next. As ever, the limit $\xi\rightarrow 0$ is a subtle one, with the ratio of energies for two adjacent bound states vanishing (or diverging) in this limit. Consequently at most only one finite energy bound-state can survive in this limit, and this agrees with the limit of the exotic bound state (discussed after \pref{eq:epsilons-sub2}) for real $\zeta$ when this limit exists. 

To summarize, this section provides general expressions for connecting bound-state and scattering observables and the RG-invariant parameters $\epscut_\star$ and $\ystar$, which reduce to the two distinct hermitian cases of \cite{Burgess2016} in the limits $\ystar \to 0$ and $\ystar \to \pm \infty$.

\subsection{Inferring RG invariants in the lab \label{sec:infer-inlab}}

We next collect explicit expressions for scattering properties, and consider how they can be inverted to find the RG invariants $\epscut_\star$ and $\ystar$. In particular we discuss how relatively exotic features of the inverse square potential, such as discrete scaling invariance, might be probed in the lab using scattering observables. Besides being of its own intrinsic interest, this also illustrates how to use PPEFT methods as a tool for parameterizing the physics of a source, and how to extract its parameters from experiments. 

We assume that each partial wave can be addressed individually, and consequently that the RG invariants for each partial wave can be measured. This is equivalent to measuring the full differential cross section ($\dd \sigma/\dd \Omega$ for $d=3$ and $\dd \sigma/\dd \theta$ for $d=2$) upon which a partial wave decomposition can be imposed. 

We consider both the absorptive and elastic cross sections as observables of choice. The former is advantageous due to its sensitivity to non-unitary physics, while the latter provides an additional probe of physics at the source, and has the obvious advantage of having a non-trivial limit for a unitary source \cite{Burgess2016,Beane2000, Coon2002, Bawin2003, Camblong2003, Braaten2004, Hammer2006}. In what follows we outline the complimentary role played by each of these scattering observables in reconstructing the RG invariants $\epscut_\star$ and $\ystar$.

Using these observables to measure the RG invariants $\ystar$ and $\epscut_\star$ provides a two-parameter fit to all physical observables that are sensitive to atom-wire interactions. This includes resonances peaks in scattering cross sections, and bound-state energies and lifetimes. While both the elastic and absorptive cross sections can be measured by performing a conventional scattering experiment, the absorptive cross section can also be extracted by measuring the decay rate of a population of trapped atoms as in \cite{Denschlag1998}. Therefore, the absorptive cross section is a somewhat more robust observable than its elastic counterpart. As before, we treat the cases where  $\zeta$ is real and imaginary separately.

\subsubsection*{When $\zeta$ is real (sub-critical) }
We begin by considering the form of the absorptive cross section for momenta satisfying  $k\ll 1/\epscut_\star$. This is guaranteed to lie within the regime of validity of our effective treatment irrespective of the size of $\epscut_\star$. 

In this limit\footnote{Since $\modA\propto (k \epscut)^\zeta$ and $\argA$ is independent of $k\epscut$ this is equivalent to working to first order in $\modA$. \label{note1}} \pref{eq:sigma-abs2} and \pref{eq:4-ela-abs-13.1} combine to give 
\begin{equation}
  k\sigma_\text{abs}\approx
  4\modA(k)\sin\frac{\pi\zeta}{2}~\sin\argA= 
  4 \sin\frac{\pi\zeta}{2}~\sin\argA
 \abs{\frac{\Gamma\qty(1-\tfrac{1}{2}\zeta)}
    {\Gamma\qty(1+\tfrac{1}{2}\zeta)}}
  \qty(\frac{k\epscut_\star}{2})^\zeta.
  \label{eq:sigma-abs-small-k}
\end{equation}
The parameters $\epscut_\star$ and $\sin\argA$ can be extracted from a plot of $\log k\sigma_\text{abs}$ vs $\log k$. Defining  $k_0$ as the momentum associated with the graph's x-intercept (i.e.\ $\log \qty[k\sigma_\text{abs}(k_0)]=0$), yields 
\begin{equation}
  \epscut_\star=2/k_0,
  \label{eq:epsilon-star-measured-sub}
\end{equation} 
while the slope of this graph, which we call  $M_\text{abs}$, allows one to extract $\sin\argA$ via the relationship
\begin{equation}\label{eq:sinargA}
  \sin\argA= \frac{M_\text{abs}}{4\zeta\sin\frac{\pi\zeta}{2}}
  \abs{\frac{\Gamma\qty(1+\tfrac{1}{2}\zeta)}
    {\Gamma\qty(1-\tfrac{1}{2}\zeta)}},\
\end{equation}  
which can be re-expressed in terms of $\ystar/\zeta$ using 
\be
  \frac{\ystar}{\zeta}=\frac{1}{\sin\argA}(1+ \cos\argA).
  \label{eq:ystar-measured-sub}
\ee

This does not yet give a unique solution since \pref{eq:sinargA} determines the magnitude, but not the sign of $\cos\argA$. The correct sign can be obtained by using --- for example --- elastic scattering data.  Using the identity $k\sigma_\text{el}=2\qty(1- \e^{-2\eta}\cos 2\delta)-k\sigma_\text{abs}$ and the small-$k\epscut$ limit\footnote{See footnote \ref{note1}. } of \pref{eq:4-ela-abs-12.2} we find 
\be
  k\sigma_\text{el}\simeq 4\modA\sin\tfrac{\pi\zeta}{2}\qty(\sin\argA\cos\Delta \pi + \cos\argA\sin\Delta\pi)+4\sin^2\tfrac{\Delta\pi}{2}-k\sigma_\text{abs}
\ee
where $\sin\argA$ is known from \pref{eq:sinargA}, while $\cos\argA$ is \emph{a-priori} unknown and determined by re-arranging the above equation to obtain 
\be
  \cos\argA=\frac{k\sigma_\text{el}+k\sigma_\text{abs} -4 \sin^2\tfrac{\Delta\pi}{2}}{4\modA\sin\tfrac{\pi\zeta}{2}\sin\Delta\pi}-\sin\argA\cot\Delta\pi,
\ee 
and when paired with \pref{eq:ystar-measured-sub} this uniquely determines\footnote{This also serves as a test of our two-parameter fit in terms of  $\epscut_\star$ and $\ystar$ as only two values are consistent with a given measurement of the absorptive cross section.}  $\ystar$. 

In the interesting case where $\epscut_\star\gg \rp$ the resonant absorption shown in \cref{fig:sub-k} is experimentally accessible. Having measured $\epscut_\star$ using the above method for $k\ll 1/\epscut_\star$ we may predict the momentum associated with resonant absorption $k_\text{res}$ shown as the vertical line in \cref{fig:sub-k}. 

To obtain an explicit formula for $k_\text{res}$ in terms of $\epscut_\star$ we seek a maximum of \pref{eq:4-ela-abs-13.1} by taking it's derivative 
with respect to $k$. This leads to the condition $\modA=1$, and by extension to the $\ystar$-independent and predictive expression
\begin{equation}
  k_\text{res}=\frac{2}{\epscut_\star}\qty[\frac{\Gamma(1+\tfrac{1}{2}\zeta)}{\Gamma(1-\tfrac{1}{2}\zeta)}]^{1/\zeta}.
  \label{eq:kres}
\end{equation}

\subsubsection*{When $\zeta$ is imaginary (super-critical)}

In this regime the rate of absorption displays oscillatory behaviour as a function of $k\epscut_\star$ (see \cref{fig:super-k}) and the value of $\ystar$ can be obtained using  $\langle k\sigma_\text{abs}\rangle_k $ averaged over different values of the momentum. This can be related to $\mathfrak{A}$ by evaluating \pref{eq:4-ela-abs-14.1} for $\cos\argA=0$ yielding
\begin{equation}
\langle k\sigma_\text{abs}\rangle_k=(1-\modA^{-2})\tanh\tfrac{\xi\pi}{2},
\end{equation} 
which can be inverted to obtain 
\begin{equation}
  \modA=\sqrt{\frac{\tanh\tfrac{\xi\pi}{2}}{\tanh\tfrac{\xi\pi}{2}-\langle k\sigma_\text{abs}\rangle_k}},
\end{equation} 
and by definition $\mathfrak{A}$ is positive and so the square root does not introduce any ambiguities. Finally, by employing \pref{eq:4-ela-abs-13} we can obtain an explicit expression for $\ystar/\xi$
\begin{equation}
  \ystar=\xi \left( \frac{1-\modA}{1+\modA} \right).
\end{equation}
In this case the value(s) of the various $\epscut^{(n)}_\star$
can be measured by identifying the peaks of the absorption, which
correspond to $\cos\argA=0$. 

In this section, we have demonstrated that in the case of the charged wire, absorptive scattering provides sufficient experimental input to explore the parameter space of the RG invariants $\ystar$ and $\epscut_\star$. The next section demonstrates the practical applicability of these results by considering a practical example that can be realized in a laboratory.

\section{Experimental protocols: past and future \label{sec:expt-prot}}
We now return to the experiment of \cite{Denschlag1998} and discuss what regions of the parameter space spanned by $\epscut_\star$ and $\ystar$ can be reasonably probed in the lab. In addition to the restriction that $\ystar<0$ (so as to ensure an absorptive source), more stringent constraints arise if one considers the observations of \cite{Denschlag1998} which imply that the wire acts as a \emph{perfect absorber}. The rate of atomic losses is well described by a completely classical theory, and given the large magnitude of applied voltages, and that the temperatures quoted are not sufficiently low to inhibit higher partial waves from scattering with the wire, this is not surprising. Indeed to obtain this limit from the quantum theory, as outlined in \cref{app:cl-abs}, the wire must act as a perfect absorber (i.e.\ one in which all waves that are infalling at the origin are absorbed), and this is equivalent to demanding that $\ystar \approx -\iu\xi$, which, as shown in \cref{fig:phase-portraits}, is a fixed point of the RG flow. As a consequence many of the interesting features discussed \S\ref{sec:RG-obs} are lost, namely: the exotic bound state, resonant scattering, and the RG-limit cycle that underlies the log-periodic absorptive cross section shown in \cref{fig:super-k}. 

Nevertheless, the phenomena outlined above can be observed in a similar experiment provided the behaviour of the compact object at the origin is modified so as to \emph{not} behave as a perfect absorber. The most obvious strategy involves altering the original set-up to include an azimuthally symmetric potential surrounding the charged wire, thereby forcing the atoms to tunnel through a barrier to access the non-unitary physics at the origin, and shielding them from its effects. A realistic implementation of this idea could be achieved---for example---by sheathing the wire in a high intensity laser beam whose frequency is blue-detuned from one of the internal level spacings of the trapped atoms, thus generating a repulsive sheathing potential due to a position dependent AC-Stark shift \cite{Grimm2000}. 

What consequences would this have for the RG invariants $\epscut_\star$ and $\ystar$? We anticipate that, although in principle $\epsilon_\star$ could also be modified by such a sheathing potential, it would primarily modify $\ystar$. This is an intuitive consequence since in the limit of an infinitely strong sheathing potential absorption would be forbidden, and we would have unitary physics at the origin. This corresponds to $\ystar=\pm \infty$ or $\ystar=0$, which is very different from the perfect absorber behaviour, $\ystar=-\xi$, which is consistent the results of the experimental results of \cite{Denschlag1998}. Therefore, for a finite laser intensity an intermediate value of $\ystar$ can be realized, and by tuning the intensity of this laser, a knob to tune $\ystar$ in the lab can be engineered. 

This then allows for the realization and observation of both exotic bound states and the RG-limit cycle behaviour's associated phenomena discussed above since one could move away from the perfect absorber fixed point of $\ystar=-\iu \xi$. We will revisit this idea in a future publication \cite{Plestid2018} and propose explicit ways to observe these effects in a lab.

\section{Summary}
\label{sec:summary}

We demonstrate how to apply PPEFTs to non-Hermitian sources by analyzing an explicit physical system that can be realized in a laboratory. Just like in the self-adjoint case, we find that the interactions between bulk fields and microscopic sources can be efficiently parameterized in terms of an action localized on a source located at the origin. This leads to a natural power counting scheme, and is amenable to RG techniques as we have shown explicitly. In particular, physical observables can be conveniently parameterized in terms of the RG invariants $\epscut_\star$ and $\ystar$, the former of which is familiar from the Hermitian case \cite{Burgess2016}. 

It is important to note that our analysis in terms of a PPEFT is not limited to the inverse square potential. Rather, this system serves as a useful toy model because it allows one to consider the centrifugal barrier as a tunable parameter, which allows probability to be sucked towards the origin, therefore increasing the system's sensitivity to point particle source. This qualitative feature is present for \emph{any} singular potential, including the $1/\vb{x}^4$ potential experienced by polarizable atoms in the presence of an ion. Likewise, PPEFT can be successfully applied to parameterize nuclear effects in precision hydrogen spectroscopy where relativistic effects can induce singular potentials \cite{Burgess:2017mhz}.

By applying this analysis to the charged wire system we provide a direct connection between the PPEFT description of atom-wire interactions and experimental observables. Furthermore, in a future publication \cite{Plestid2018} we plan to present an explicit experimental proposal to realize the consequences of the re-normalization procedure describe above in a lab. Such a proposal, if realized, could serve as a testing ground for the progress of the past two decades concerning the theoretical treatment of the inverse square potential \cite{Burgess2016, Beane2000, Coon2002, Bawin2003, Camblong2003, Braaten2004, Hammer2006, Kaplan2009}, and in particular the anomalous breaking of scale invariance.  Signatures include a single exotic bound state,  an Efimov like tower of bound-states, or a log-periodic absorptive cross section as a function of momentum.


\section*{Acknowledgements}
 We would like to thank Peter Hayman for his detailed criticism of early versions of this work, especially regarding the near-source expansions of the mode functions. We also thank L\'aszl\'o Z\'alviari, Markus Rummel, Marvin Holten, Sung-Sik Lee, Yvan Castin, and Felix Werner for useful discussions. This work was partially supported by funds from the Natural Sciences and Engineering Research Council (NSERC) of Canada. Research at the Perimeter Institute is supported in part by the Government of Canada through NSERC and by the Province of Ontario through MRI. 

\appendix

\section{Absorptive scattering}
\label{app:abs-sc}

This appendix summarizes the quantum mechanical treatment of scattering with an absorptive component (see e.g. \cite{Landau1977} \S142). 

\subsection{Phase shifts}

In unitary Schr\"odinger scattering one assumes boundary conditions at spatial infinity that correspond to the superposition of an incoming plane wave (of known amplitude, moving along the $z$-axis say) plus and outgoing scattered spherical wave of initially unknown amplitude. One uses boundary conditions at the origin (plus solves for Schr\"odinger evolution through any potential that is present) and thereby determines the amplitude of the outgoing wave. 

The same logic applies in the case of an absorptive scattering centre, with the only difference being that Im$\,h \ne 0$ implies the boundary condition applied at the origin is not unitary inasmuch as there is a nonzero flux of probability --- {\em c.f.}~eq.~\pref{netprobloss} --- flowing into the origin through the surface at $r = \epscut$. Because probability is conserved everywhere {\em except} at the origin, the same amount of net flux also flows through a large sphere with radius $r \to\infty$. 

For large $r$ the asymptotic form of the mode functions can be evaluated in the standard way, leading in three dimensions to
\be \label{3dasympt}
 \psi_\ell = A_\ell \,h_\ell^{(1)} - B_\ell \,h_\ell^{(2)}  \simeq \frac{1}{r} \qty[A_\ell \,\e^{\iu(kr-\ell\pi/2)}+B_\ell \,\e^{-\iu(kr-\ell\pi/2)} ] \,,
\ee
where $h_{\ell}^{(1,2)}$ are spherical Hankel functions. The corresponding result in two dimensions is instead
\be\label{2dasympt}
  \psi_\ell= A_\ell \,H_\ell^{(1)}+B_\ell \,H_\ell^{(2)} \simeq  \frac{1}{\sqrt{r}}
\qty[A_\ell \,\e^{\iu(kr-\ell\pi/2-\pi/4)}+B_\ell \,\e^{-\iu(kr-\ell\pi/2-\pi/4)} ] \,,
\ee
with $H_\ell^{(1,2)}$ denoting the ordinary Hankel functions. 

In both cases $A_\ell$ and $B_\ell$ are integration constants, whose ratio is fixed by the boundary condition at $r = \epscut$ (see eq.\ \ref{eq:Spbcmode}). The $S$-matrix for each partial wave is defined in the usual way \cite{Landau1977,Sakurai1988} in terms of $A_\ell/B_\ell$, giving
\begin{align}
  S_\ell := \e^{2\iu\gamma_\ell} 
  &=-\frac{A_\ell}{B_\ell} \quad\; \hbox{if $d=3$} 
      \label{eq:4-ela-abs-3.1}\\
  &=\iu~\frac{A_\ell}{B_\ell}\quad\quad \hbox{if $d=2$} \,.
    \label{eq:4-ela-abs-3.2}
\end{align}
The factors of  $\iu$ and $-1$ in these expressions are related to the partial-wave expansion of the incoming plane wave in the appropriate dimension. The corresponding partial-wave $T$-matrix element then is  
\begin{equation}
  T_\ell = \frac{1}{2\pi \iu}(S_\ell -1) \,,
  \label{eq:4-ela-abs-4}
\end{equation}
where we use the normalization conventions of \cite{Sakurai1988}.

Expressing the integrated radial probability flux at fixed radius in terms of partial waves gives
\begin{equation}
  \oint J_r \, r^{d-1} \dd \Sigma = \sum_\ell \frac{r^{d-1}}{2\iu m}
  \qty[\psi^*_\ell \dv{r} \psi_\ell - \psi_\ell \dv{r}\psi^*_\ell ] 
  \label{eq:4-ela-abs-1}
\end{equation}
which uses the orthogonality of the spherical/circular  harmonics. Inserting the asymptotic forms \pref{3dasympt} and \pref{2dasympt} into \pref{eq:4-ela-abs-1}, it follows that 
\begin{align}
  \oint J_r\, r^2 \dd \Sigma
  \underset{r\rightarrow\infty}{\to}& 
  \frac{k}{m}
  \sum_\ell \qty[\abs{A_\ell}^2-\abs{B_\ell}^2] \qquad \hbox{if $d=3$}\,.
    \label{eq:4-ela-abs-2.1}\\
  \oint J_r\, r \dd \theta 
  \underset{r\rightarrow\infty}{\to}&
  \frac{k}{m}
  \sum_\ell \qty[\abs{A_\ell}^2-\abs{B_\ell}^2] \qquad \hbox{if $d=2$} \,.
    \label{eq:4-ela-abs-2.2}
\end{align}
For unitary scattering these integrals vanish, which shows why $S_\ell$ is in this case a pure phase and $\gamma_\ell$ is real. But these integrals do not vanish for absorptive scattering --- being given instead by \pref{netprobloss} --- so we regard the phase shift as complex, denoting its real and imaginary parts by 
\be
   \gamma_\ell := \delta_\ell +\iu\eta_\ell \,.
\ee 

\subsection{Cross sections}

The elastic cross section is defined using the $S$-matrix just as it would have been if the scattering had been unitary, leading to the following expressions for the $\ell$th partial wave (see \cite{Landau1977, Sakurai1988} for $d=3$  and \cite{Adhikari1986} for $d=2$)
\begin{align}
  \sigma^{(\ell)}_{\textrm{el}}&= \frac{\pi}{k^2}(2\ell +1)
  \abs{1-S_\ell}^2
  \qquad \hbox{if $d=3$}    \label{eq:4-ela-abs-6.1}\\
  \sigma^{(\ell)}_{\textrm{el}}&=~ \frac{1}{k} \abs{1-S_\ell}^2
  \quad\qquad\qquad\;\; \hbox{if $d=2$} \label{eq:4-ela-abs-6.2}\,.
\end{align}
Physically this measures the flux due to the outgoing spherical wave, normalized by the flux of the incident plane wave. It reduces to the standard result for unitary scattering result, and extends it to when absorption is present. Re-expressed in terms of the complex phase shifts $\gamma_\ell$ these become
\begin{align}
  \sigma^{(\ell)}_{\textrm{el}}
  &=\frac{\pi}{k^2}(2\ell +1)
  \Bigl[1+ \e^{-4\eta_\ell} - 2 \e^{-2\eta_\ell}  \cos 2\delta_\ell\Bigr]
  \qquad \hbox{if $d=3$}    \label{eq:4-ela-abs-7.1}\\
  \sigma^{(\ell)}_{\textrm{el}}
  &=\frac{1}{k} \Bigl[1+ \e^{-4\eta_\ell} - 2\e^{-2\eta_\ell}\cos 2\delta_\ell\Bigr]
  \qquad\qquad\quad\;\;\; \hbox{if $d=2$}\label{eq:4-ela-abs-7.2}.
\end{align}

By contrast, the absorptive cross section is defined as the ratio of the \emph{total} inward
probability flux $\vb{J}\cdot(-\vb{\hat{r}})$ normalized by the incident flux. This is given, up to the same pre-factors appearing in the elastic cross section, by $1-\abs{S_\ell}^2 = 1 - \e^{-4\eta_\ell}$ which parameterizes the difference in the amplitude of the incoming and outgoing spherical waves. Therefore \cite{Landau1977}
\begin{align}
  \sigma^{(\ell)}_{\textrm{abs}} &= \frac{\pi}{k^2}(2\ell +1)
  \Bigl( 1- \e^{-4\eta_\ell}\Bigr)  \qquad \hbox{if $d=3$}
    \label{eq:4-ela-abs-8.1}\\
   \hbox{and} \quad \sigma^{(\ell)}_{\textrm{abs}} &= \frac{1}{k} \Bigl( 1- \e^{-4\eta_\ell} \Bigr) \qquad\qquad\quad\;\;\; \hbox{if $d=2$} \,.
    \label{eq:4-ela-abs-8.2}
\end{align}
Eqs.\ (\ref{eq:4-ela-abs-7.1}--\ref{eq:4-ela-abs-8.2}) are the required expression for the cross sections, and from them we see that all cross sections can be computed given expressions for the two quantities $\e^{-4\eta_\ell}$  and $\e^{-2\eta_\ell}\cos 2\delta_\ell$.

\subsection{Classical fall to the center \label{app:cl-abs}}
In this section we follow \cite{Alliluev1972} and show how the classical absorption cross section emerges from the above formulae, under the assumption that the quantum process is a perfect absorber. 

Classical motion in the inverse square potential leads to a fall to the center in finite time for any trajectory with angular momentum less than a critical value $\ell_c=2mg$  \cite{Landau1982}. In a scattering experiment at fixed energy $2mE = k^2$ this implies a critical impact parameter $b_c = \ell_c/k$, and all trajectories with impact parameter smaller than this inevitably fall to the center. This leads to the following classical absorption cross section \cite{Landau1982}
\bea
   \sigma_\text{abs}&=&\pi b_c^2 \qquad \hbox{if $d=3$} \nn\\
   \sigma_\text{abs}&=& 2 b_c^2 \qquad\; \hbox{if $d=2$} \,.
   \label{eq:cl-sigma}
\eea

These same results can be obtained from eqs.\pref{eq:4-ela-abs-8.1} and \pref{eq:4-ela-abs-8.2} in the limit that $\ell_c\gg 1$, provided the `perfect absorber' boundary condition, $C_+/C_-=0$, is imposed, corresponding to the fixed point with $\ystar=-\xi$ in the regime where $\zeta = \iu \xi$ is imaginary. As is shown in the main text --- {\em c.f.} the discussion around \pref{perfectabsform} --- this choice implies $4\eta_\ell = \xi\pi $.

One can then consider the sum of the partial wave cross-sections
\begin{equation}
  \sigma_\text{abs}=\sum_\ell\sigma^{(\ell)}_{\text{abs}}\approx 
  \int_0^\infty \sigma_\text{abs}(\ell)\dd \ell = 
  \begin{cases}
    \frac{\pi}{k^2} \int_0^{\ell_c }
  (2\ell+1) \qty(1-\e^{-4\eta(\ell)})\dd \ell & \text{for $d=3$}\\
    \frac{1}{k} \int_0^{\ell_c }
    2 \qty(1-\e^{-4\eta(\ell)})\dd \ell & \text{for $d=2$}\\
  \end{cases}
  \label{eq:sigma-abs-integral}
\end{equation}
where the factor of 2 in $d=2$ accounts for both positive and negative $\ell$.
The upper bound $\ell_c$ is a consequence of the fact that for $\ell>\ell_c$, 
$\zeta$ is real and absorption ``turns off''.

For the perfect absorber, when $\ell<\ell_c$ the value of $\zeta=\iu\xi$ is imaginary and $4\eta_\ell=\pi \xi(\ell)$ (see \pref{eq:4-ela-abs-14.1} in the limit that $\modA\rightarrow\infty$). In any dimension therefore 
\be
 4 \eta =  \xi  \pi = 2\pi \sqrt{w_d^2-\varpi_d(\ell)} \quad \text{(perfect absorber)}\,,
\ee
where $w_d^2 := 2mg - \frac14 (d-2)^2$. Furthermore, for $d = 3$ we have $\exd \varpi_3 = (2\ell + 1) \exd \ell$ so the desired result can be rewritten as an elementary integral
\be \label{integralresult}
 \cI := \int_0^{w_3^2} \exd \varpi_3 \Bigl( 1 - \e^{-2\pi \sqrt{w_3^2 - \varpi_3}} \Bigr) 
 = w_3^2 - \frac{1}{2\pi^2} \Bigl[ 1 - (1 + 2\pi w_3) \e^{-2\pi w_3} \Bigr] \,.
\ee
Therefore $\cI \simeq w_3^2$ whenever $w_3 \gg 1$. Since for any dimension $\varpi_d(\ell)\sim \ell^2$ at large $\ell$ and since in this same limit $\ell_c \simeq w_d$, it follows that for both $d = 3$ we have $\cI \simeq \ell_c^2$ and so
\be 
  \sigma_\text{abs} \approx \frac{\pi \ell_c^2}{k^2} = \pi b_c^2 \quad \hbox{(if $d=3$)} \,,
\ee
where $b_c$ is the classical impact parameter, in agreement with the classical result  \pref{eq:cl-sigma}.

Although the integral cannot be similarly reduced to elementary quadratures when $d=2$, the main lesson of \pref{integralresult} is that for large $w_d$ the integrand is well-approximated over most of the integration range by dropping the exponential term altogether. Evaluating the integral using this same approximation for $d = 2$ then leads to the expression
\be 
  \sigma_\text{abs} \approx \frac{2 \ell_c}{k} =2 b_c \quad \hbox{(if $d=2$)} \,,
\ee
again in agreement with the classical geometrical cross section.

\section{Parameterization of the RG flow \label{app:RG}}

The $\epscut$ dependence of the PPEFT's coupling $h(\epscut)$, is defined via  \pref{psiellbc} which is repeated for context
\begin{equation} 
  \begin{split}
 \frac{2mh}{\OmegaD \epscut^{d-1}} &=  \Bigl( \partial_r \ln \psi \Bigr)_{r=\epscut} = \left(  \frac{C_{+} \psi'_{+} + C_{-} \psi'_{-}}{C_{+} \psi_{+} + C_{-} \psi_{-}} \right)_{r=\epscut} \nn\\
 &:=\frac{1}{2\epscut}(2-d+\hLambda)\\
 &\simeq  \frac{1}{\epscut} \left[\frac{(B_{+}/B_{-}) s_+(k \epscut)^{s_+-s_-} + (C_{-}/C_{+}) s_-}{(B_{+}/B_{-})(k\epscut)^{s_+-s_-} + (C_{-}/C_{+}) })\right],
 \end{split}\,,
\end{equation}
where we have used $s_\pm=(d-2\pm\zeta)/2$, and introduced a new variable $\hLambda$ for convenience, which is defined exactly in terms of the bulk mode functions. This appendix is dedicated to the validity of the approximation made in the second line of \pref{psiellbc}, as well as the complimentary topic of the relationship between $\hLambda(\epscut)$ (or equivalently $h(\epscut)$) and $\hlambda(\epscut)$ as defined in \pref{lambdadef} and \pref{Rdef}
\begin{equation} 
  \hlambda:=\zeta \qty[\frac{1-R}{1+R}]\quad \text{with} \quad R(\epscut):=\frac{C_-}{C_+}(2 \iu k\epscut)^{-\zeta}.
\end{equation}

\subsection{Asymptotic expansions}

\subsubsection{The case of $\textrm{Re} ~\zeta<2$}

The asymptotics for $\psi_\pm$  can be found using either the confluent hypergeometric (or Bessel) representation presented in \pref{psielldef}. The series expansion about the origin is given by 
\be\label{psiasym}
\psi_\pm(z)=
   (2z)^{\tfrac{1}{2}(d-2 \pm\zeta)}\sum_{k=0}^\infty \frac{1}{k!}\frac{\Gamma\qty(1\pm \frac{\zeta}{2})}{\Gamma(1\pm\frac{\zeta}{2}+k)}\qty(\tfrac{1}{2}z)^{2k}.
\ee
We are ultimately interested in studying
\begin{equation}
  \frac{1}{2\epscut}(2-d+\hLambda)=\frac{\psi'_+ + (C_-/C_+)\psi'_-}{\psi_+ +(C_-/C_+)\psi_-}
\end{equation}
and from \cref{psiasym} we see that keeping only the leading order (LO) contribution of $\psi_-$ and $\psi_+$ captures the LO, and next to LO (NLO) contributions of both the numerator and denominator, \emph{if and only if} $\textrm{Re}~\zeta<2$. When this condition is not satisfied, sub-leading corrections to $\psi_-$ ($\psi'_-$) will be parametrically larger than the LO contribution from $\psi_+$ ($\psi'_+$). We will discuss this situation in the subsequent section.

Returning to the case of $\textrm{Re}~\zeta<2$ we recover the approximation made in \pref{psiellbc} with $B_+/B_-= (2i)^{2\zeta}$, and find that $\hLambda(\epscut)$ and $\hlambda(\epscut)$ are degenerate in the $k\epscut\rightarrow 0$ limit
\begin{equation}
  \begin{split}
   \frac{1}{2\epscut}(d-2+\hLambda)&\sim \frac{1}{2\epscut}
   \frac{ (d-2+\zeta) (2k\epscut)^\zeta\qty[1+\order{(k\epscut)^2}] + (d-2-\zeta)(C_-/C_+)\qty[1+\order{(k\epscut)^2}]}
        {(2k\epscut)^\zeta\qty[1+\order{(k\epscut)^2}] + (C_-/C_+)\qty[1+\order{(k\epscut)^2}] }\\
      &\sim\frac{1}{2\epscut}\qty((d-2) \frac{1+R(\epscut)}{1+R(\epscut)} + 
      \zeta\frac{1-R(\epscut)}{1+R(\epscut)}) + \order{(k\epscut)^2}\\
      &\sim \frac{1}{2\epscut}\qty(2-d +\hlambda) +\order{(k\epscut)^2}.
  \end{split}
  \label{Lambda-asym}
\end{equation}

\subsubsection{The case of $\textrm{Re} ~\zeta\geq2$ \label{sec:intermediate}}
The approach of the previous section can be applied for larger values of $\zeta$ truncating the expansions multiplied by $(C_-/C_+)$ at $\order{(k\epscut)^{2n}}$ where $n$ is the largest integer satisfying $2n<\zeta$.  Explicitly the corrected asymptotic form is obtained from \pref{Lambda-asym} by the following substitutions 
\begin{equation}\label{Lambda-subs}
  \begin{split}
    (d-2-\zeta)\frac{C_-}{C_+}\qty[1+\order{(k\epscut)^2}] &\rightarrow \frac{C_-}{C_+}\qty[\sum_{k=0}^n \frac{4k+2s_-}{k!}\frac{\Gamma\qty(1\pm \frac{\zeta}{2})}{\Gamma(1\pm\frac{\zeta}{2}+k)}\qty(\tfrac{1}{2} k\epscut)^{2k+s_-}]\\
    \frac{C_-}{C_+}\qty[1+\order{(k\epscut)^2}] &\rightarrow \frac{C_-}{C_+}\qty[\sum_{k=0}^n \frac{1}{k!}\frac{\Gamma\qty(1\pm \frac{\zeta}{2})}{\Gamma(1\pm\frac{\zeta}{2}+k)}\qty(\tfrac{1}{2}k \epscut)^{2k+s_-}],
  \end{split}
\end{equation}
for the numerator and denominator respectively, and where $s_-=\tfrac{1}{2}(d-2-\zeta)$. 

\subsubsection{The limit of $\zeta\rightarrow\infty$ \label{app:surprise} }
Although at intermediate values of $\zeta$ multiple terms in the above expansion must be kept, we will show that for very large $\zeta$ the small $k\epscut$ form of the mode functions provides a good approximation. We make use of asymptotic expansions for fixed $k\epscut$ in the limit that $\zeta\rightarrow \infty$, in particular \cite{NIST2010} 
\begin{align}
  I_{-\nu}(z)&=I_\nu (z) + \frac{2\sin\nu\pi}{\pi} K_\nu(z)\\
  I_{\nu}(z)&\sim \frac{1}{\sqrt{2\pi\nu}}\qty(\frac{\e z}{2\nu})^\nu 
  \quad \text{as} ~\nu\rightarrow\infty~ \text{for fixed}~ z\neq0\\
  K_{\nu}(z)&\sim \sqrt{\frac{\pi}{2\nu}}\qty(\frac{\e z}{2\nu})^{-\nu} 
  \quad \text{as} ~\nu\rightarrow\infty~ \text{for fixed}~ z\neq0\\
\end{align}
which upon using \pref{psielldef} implies that for fixed $z\neq0$ we have
\begin{align}\label{asymhizeta}
  \psi_+(k\epscut)&\sim (2\iu k\epscut)^{\tfrac{1}{2}(d-2)}\frac{1}{\sqrt{\pi\zeta}}\qty(\frac{\e\iu k\epscut}{\zeta})^{\tfrac{1}{2}\zeta}\\
  \psi_-(k\epscut)&\sim (2\iu k\epscut)^{\tfrac{1}{2}(d-2)}\frac{2\sin\tfrac{\pi\zeta}{2} }{\sqrt{\pi\zeta}} \qty(\frac{\e \iu k\epscut}{\zeta})^{-\tfrac{1}{2}\zeta}
\end{align}
which has the form of \pref{psiellbc}, but with modified coefficients $B_\pm$. We note, as is typical when dealing with modified Bessel (or confluent hypergeometric) functions, integer values of $\nu$ ($\zeta$) are treated using a limiting procedure. 

It is worth emphasizing that the limiting form shown in \cref{asymhizeta} takes $k\epscut /\zeta$ as an argument; this behaviour is peristent in the full asymptotic series (i.e.\ $\psi_\pm(\zeta z)\sim \sum_\zeta f_\zeta(z)$ as implied by \S10.41 of \cite{NIST2010}). Consequently the function is stretched by a factor of $\zeta$, and so the small-$k\epscut$ expansion provides a faithful approximation provided  $k\epscut\lsim \zeta/e$ rather than the naive expectation of $k\epscut\lsim1$. This implies that the small $k\epscut$ approximation's regime of validity as a function of $k\epscut_\star$ expands as $\zeta\rightarrow \infty$, and is actually valid at large $\zeta$ as claimed at the begining of this section. 

In summary, we have found that $\hlambda$ and $\hLambda$ are simply related in the near-source regime for both $\zeta\gg 2$ and $\zeta<2$. The former region generically give small values of $\epscut_\star$, while in the latter case larger values are more likely to occur. We have also showed how to systematically correct the expansion with $\zeta$ fixed for $\zeta\geq 2$. Finally we note that while $\hlambda$ and $\hLambda$ behave similarly in the near-source regime, this is not true for $\epscut\gsim 1$, and this is most striking when $k\epscut_\star\gsim 1$. We note, however, that for fixed $\epscut_\star$ in the limit that $k\rightarrow0$ the near-source expansion becomes reliable.  


\subsection{Relating $\hLambda$ to $\hlambda$ \label{app:Lamlamrel}}
In this section we provide explicit expressions for both $\hlambda(\hLambda)$ and $\hLambda(\hlambda)$ demonstrating, as claimed in the main text, that the mapping is bijective. We begin by using \pref{Rdef} to re-express $C_-/C_+$ in terms of $\hlambda$. This expression can be substituted into the function $\hLambda(C_-/C_+)$ \pref{psiellbc} to obtain
\begin{equation}
  \frac{1}{2\epscut}(2-d+\hLambda)= 
  \frac{\partial_r \psi_+ + \frac{\zeta-\hlambda}{\zeta+\hlambda}
    (2\iu k\epscut)^\zeta \partial_r \psi_-}
       {\psi_+ + \frac{\zeta-\hlambda}{\zeta+\hlambda}
         (2\iu k\epscut)^\zeta\psi_- }
\end{equation}
with $\psi_\pm$ given by \pref{psielldef}. This gives $\hLambda$ explicitly as a function of $\hlambda$. 

Conversely to find $\hlambda$ as a function of $\hLambda$ we may first invert 
\pref{psielldef} to find $C_-/C_+$ in terms of $\hLambda$  
\begin{equation}
  \frac{C_-}{C_+}=\frac{ \psi'_+ -\tfrac{1}{2\epscut}(2-d+\hLambda)\psi_+}{\frac{1}{2\epscut}(2-d+\hLambda)\psi_- -\psi'_+}
\end{equation}
which then gives $\hlambda$ directly upon use of \pref{lambdadef} and \pref{Rdef}.  

This shows that the mapping between $\hLambda$ and $\hlambda$ is bijective for all $\epscut$, and this guarantees the RG flow for both quantities will have the same topology. 
\subsection{RG flow in the infrared: interpretation of $\hlambda$ vs $\hLambda$}
The flow of $\hlambda$ is essentially governed by extrapolating the behaviour of the near-source form of the mode functions $\psi_{\pm}$ to arbitrarily large distances. In contrast, the function $\hLambda(\epscut)$ tracks the evolution of the mode functions deep into the bulk. Given that the interactions in $S_p$ are by definiton local, only the near-source behaviour of the mode functions should be relevant when considering interactions with the source. In what follows, we argue that the consequence of this observation is that it is the behaviour of $\hlambda$ under the RG, as opposed to $\hLambda$, that identifies the physical length scale introduced by the source. This length scale manifests itself in the scattering length and bound state energies, and can consequently be regarded as physical.
\begin{figure}[!h]
  \centering
  \begin{subfloat}[][$\zeta=0.1$]%
  {  \includegraphics[clip,width=0.45\linewidth]{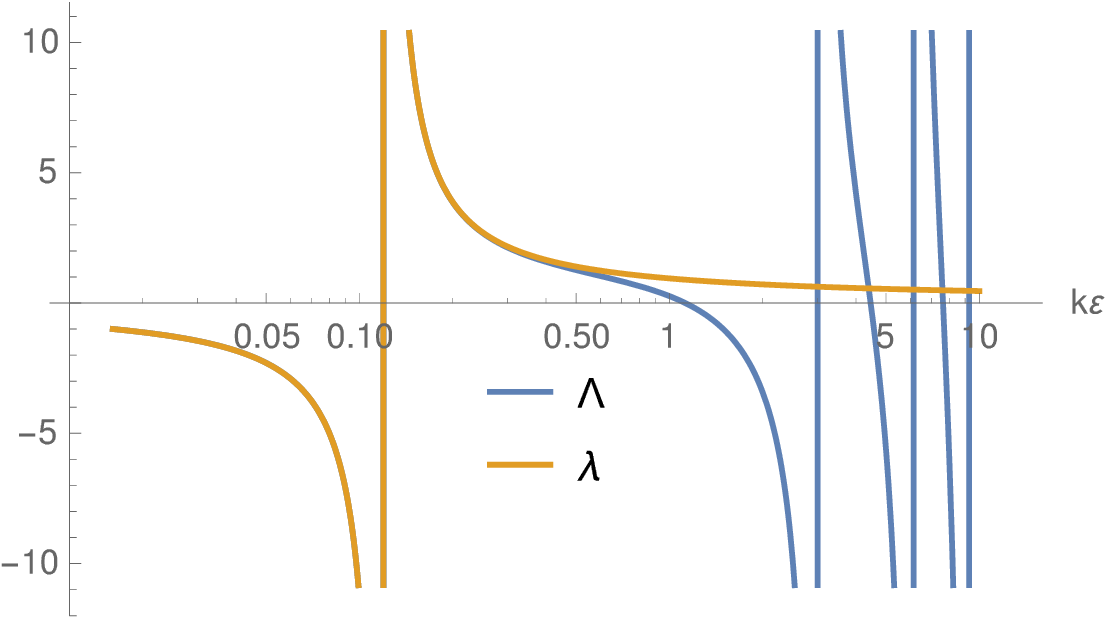}    \label{fig:LL1}      }
  \end{subfloat}%
 \begin{subfloat}[][$\zeta=2.15$]%
   {    \includegraphics[clip,width=0.45\linewidth]{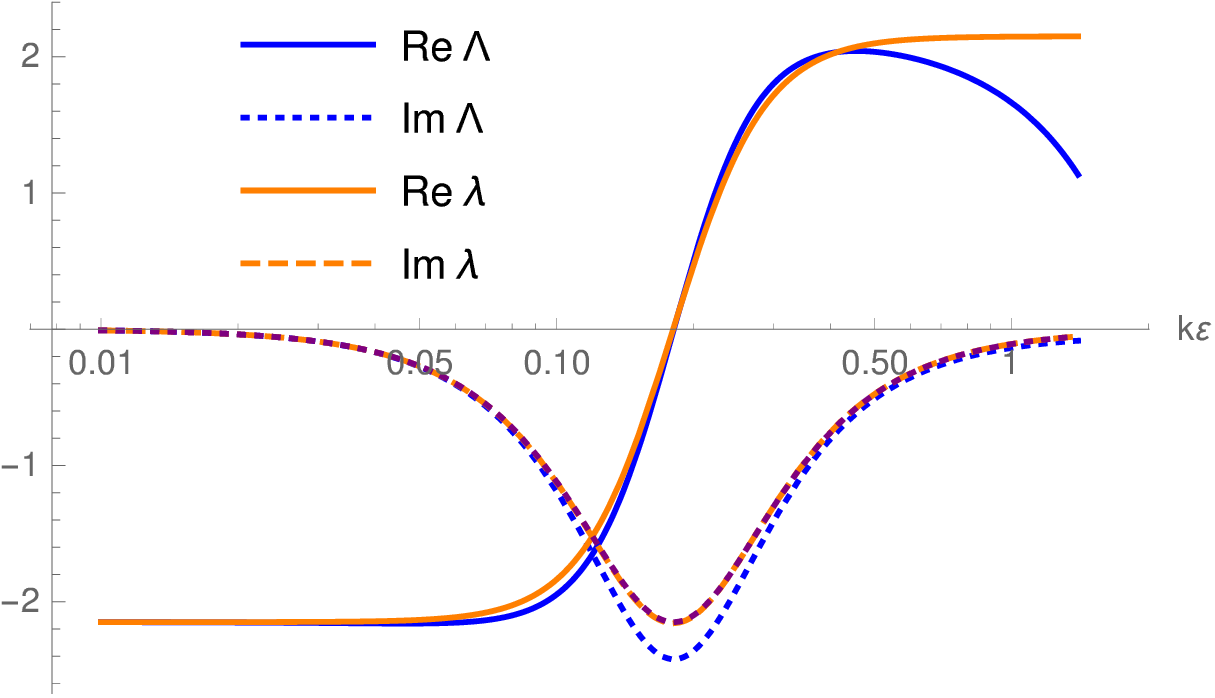}    
     \label{fig:LL2}  } %
 \end{subfloat}
 \caption{ Behaviour of $\hLambda$ (defined in terms of the exact mode functions), and $\hlambda$ (extrapolated near-source behaviour) as a function of $k\epscut$.  A comparison for small $\zeta$ is shown in \protect\subref{fig:LL1}, where the consequence of the oscillatory behaviour of the exact mode functions can be seen in $\hLambda$ as a $\tan(k\epscut)$ dependence as $k\epscut\rightarrow\infty$. The case of intermediate $\zeta$ is shown  in \protect\subref{fig:LL2} with parameters $\zeta=2.15$ and $\ystar=-\zeta$ corresponding to the resonant absorption shown in \cref{fig:sub-k}. We see good agreement between $\hlambda$ and $\hLambda$ at the resonant momentum $\epscut \approx 0.5/k_\text{res}$ in spite of the fact that $\zeta\approx 2.15$ and $\epscut k_\text{res}\sim \order{1}$. The initial condition for both figures is $\epscut_0=10^{-3}/k$. In contrast  $\epscut_\star\approx 10^{-1}/k$ which is approximately two orders of magnitude larger, and so furnishes an example in which the scattering length is anomalously large in comparison to the size of the source.  \label{fig:LL} }
\end{figure}
The RGE governing the evolution of $\hlambda$ can be understood by considering the parametric size of the contributions from the two power-law solutions $\psi_+\propto(k\epscut)^\zeta$ and $\psi_-\propto(k\epscut)^{-\zeta}$. Fixing $C_-/C_+$ at small $\epscut$ defines an initial value of $\hlambda$, which flows along the trajectories outlined in \cref{fig:phase-portraits} moving from the UV fixed point at $\hlambda=-\zeta$ to the IR fixed point at $\hlambda=\zeta$; the former corresponds to $C_-/C_+=\infty$ while the latter is equivalent to $C_-/C_+=0$. This fixed point structure is a consequence of the monotonic nature of the power-law solutions outlined above. 

In contrast the flow of $\hLambda$ is controlled by the exact mode-functions whose bulk behaviour is non-monotonic, transitioning from power-law behaviour at the origin, to oscillatory behaviour deep in the bulk. This difference in behaviour can be seen by comparing the evolution of $\hLambda$ and $\hlambda$ as is illustrated in \cref{fig:LL1} where the flow is degenerate until $k\epscut\sim\order{1}$ and then a $\tan(k\epscut)$ like structure emerges. This change in behaviour does not, however, play a role in the interactions with the source. We should therefore not expect the behaviour of $\hLambda$ and large $k\epscut$ to serve as a useful diagnostic of interactions with the compact object (i.e.\ atom-wire interactions in our explicit charged-wire example). 

This statement is further vindicated by the explicit expressions for bulk observables, such as scattering lengths, and bound state energies, whose length scale is set naturally by $\epscut_\star$.  As emphasized above, this is obtained by extrapolating the near-source behaviour of the mode functions. Finally, we note that although there may exist accessible momenta $k$ such that $\hLambda(k\epscut)$ and $\hlambda(k\epscut)$ are very different, for fixed $\epscut$ in the limit of $k\rightarrow 0$ these two functions necessarily become degenerate.

\providecommand{\href}[2]{#2}\begingroup\raggedright\endgroup

\end{document}